\documentclass[aps,preprint,tightenlines]{revtex4}%
\usepackage{amsmath}%
\usepackage{amsfonts}%
\usepackage{amssymb}%
\usepackage{graphicx}

\begin{document}

\title{Ground state and optical conductivity of interacting polarons in a quantum dot}
\author{S. N. Klimin$^{\ast}$, V. M. Fomin$^{\ast,\ast\ast}$, F. Brosens, and J. T.
Devreese$^{\ast\ast}$}
\affiliation{Theoretische Fysica van de Vaste Stoffen (TFVS), Universiteit Antwerpen,
B-2610 Antwerpen, Belgium}

\begin{abstract}
The ground-state energy, the addition energies and the optical absorption
spectra are derived for interacting polarons in parabolic quantum dots in
three and two dimensions. A path integral formalism for identical particles is
used in order to take into account the fermion statistics. The approach is
applied to both closed-shell and open-shell systems of interacting polarons.
Using a generalization of the Jensen-Feynman variational principle, the
ground-state energy of a confined $N$-polaron system is analyzed as a function
of $N$ and of the electron-phonon coupling constant $\alpha$. As distinct from
the few-electron systems without the electron-phonon interaction, \emph{three}
types of spin polarization are possible for the ground state of the
few-polaron systems: (i) a spin-polarized state, (ii) a state where the spin
is determined by Hund's rule, (iii) a state with the minimal possible spin. A
transition from a state fulfilling Hund's rule, to a spin-polarized state
occurs when decreasing the electron density. In the strong-coupling limit, the
system of interacting polarons turns into a state with the minimal possible
spin. These transitions should be experimentally observable in the optical
absorption spectra of quantum dots.

\end{abstract}
\date{May 21, 2004}
\startpage{1}
\endpage{ }
\maketitle

\section{Introduction}

Many-electron states in quantum dots have been theoretically investigated by
various approaches, e.g., the Hartree-Fock method
\cite{PRL82-5325,PRB59-13036,PRB61-1971}, the density-functional theory
\cite{PRB57-9035,PRB59-4604,EPJ9-105,PRB61-5202,PRB63-045317,P2000-1,P2000-2},
the quantum Monte Carlo simulation \cite{PRL82-3320}, the variational Monte
Carlo method and the Pad\'{e} approximation \cite{P99-1,P99-2}, a numerical
diagonalization of the Hamiltonian in a finite-dimensional basis
\cite{PRB62-8108}. The electron-phonon interaction was not taken into account
in these investigations, although it can contribute significantly to both
equilibrium and non-equilibrium properties of quantum dots. For instance, the
effects due to the electron-phonon interaction play a key role in the optical
spectra of some quantum dots (see Ref. \cite{PRB57-2415} and references
therein). Some characteristics revealed {in the mid-infrared region of} the
experimentally observed optical absorption spectra of high-$T_{c}$ cuprates
\cite{Genzel,Thomas,CalvaniSSC,Calvani,Lupi98} were assigned \cite{DT98} to
the polaron optical absorption at intermediate values of the electron-phonon
coupling constant $\alpha$. The infrared optical absorption band in the
neodymium-cerium cuprate Nd$_{2-x}$Ce$_{x}$CuO$_{4-y}$ (NCCO), which was
studied experimentally \cite{Lupi99} as a function of the electron density,
has been associated with the polaron optical absorption. Also some
peculiarities of the infrared optical absorption spectra of cuprates and
Pr$_{2}$NiO$_{4.22}$ were interpreted in terms of a mixture of large and small
polarons or bipolarons \cite{Emin93,Eagles}.

The theory of the optical conductivity of arbitrary-coupling single polarons
has been developed in Refs. \cite{DSG,Green} (see also the review
\cite{Polarons} and references therein) within the memory-function method
based on the path-integral formalism \cite{Feynman55}. Recently, the optical
absorption of a gas of interacting polarons at weak coupling was investigated
\cite{TD1,TD2} on the basis of a variational many-particle approach
\cite{LBD}. The optical absorption of interacting polarons in bulk
semiconductors at arbitrary coupling strength has been treated in Ref.
\cite{Iadonisi} within the random-phase approximation, using variational
parameters obtained from Feynman's \emph{single-polaron} model
\cite{Feynman55}, but this treatment does not seem to be a self-consistent
approach to the many-polaron problem.

In contrast to the polaron mechanism of optical absorption in bulk (see Refs.
\cite{Polarons,Devreese2003} and references therein), the polaron optical
absorption in quantum dots, to the best of our knowledge, has not yet been
widely studied. In order to investigate the ground state and the optical
response of a system of interacting polarons in a quantum dot, it is crucial
to take into account the fact that the system contains a \emph{finite} number
of identical particles (electrons or holes). Indeed, the thermodynamic
properties of systems with a finite number of identical particles might
substantially deviate from those obtained within the grand-canonical formalism
(see, e.g., \cite{PRE97,SSC99}). The variational path-integral method for
identical particles \cite{PRE97,SSC99,PRE96,Note} provides a useful tool for
investigating interacting quantum many-body systems with a fixed (few or many)
number of $N$ particles. An outline of the method is given in Ref.
\cite{SSC114-305}, where we sketched the calculation of the ground-state
energy of a fixed number of interacting polarons, which form a closed-shell
system in a quantum dot.

In the present paper the calculation of the ground state and of the optical
conductivity is performed for both closed-shell and open-shell systems of
interacting polarons in a quantum dot. In Section \ref{sec:Freeenergy}, we
derive an upper bound to the free energy of a finite number of interacting
polarons confined in a parabolic quantum dot in three dimensions and in two
dimensions. In Section \ref{sec:groundstate} we discuss the numerical results
for the ground-state energy and for the addition energy of this system. In
Section \ref{sec:optconductivity}, the optical conductivity of interacting
polarons in a quantum dot is derived on the basis of the memory-function
method. The numerical results for the optical conductivity are discussed in
Section \ref{sec:resultsconductivity}. The last section \ref{sec:conclusions}
contains conclusions.

\section{The partition function and the free energy of a many-polaron system}

\label{sec:Freeenergy}

\subsection{Interacting polarons in a quantum dot}

We consider a system of $N$ electrons with mutual Coulomb repulsion and
interacting with the lattice vibrations. The system is assumed to be confined
by a parabolic potential characterized by the frequency parameter $\Omega_{0}%
$. The total number of electrons is represented as $N=\sum_{\sigma}N_{\sigma
},$ where $N_{\sigma}$ is the number of electrons with the spin projection
$\sigma=\pm1/2$. The electron 3D (2D) coordinates are denoted by
$\mathbf{x}_{j,\sigma}$ with $j=1,\cdots,N_{\sigma}.$ The bulk phonons
(characterized by 3D wave vectors $\mathbf{q}$ and frequencies $\omega
_{\mathbf{q}}$) are described by the complex coordinates $Q_{\mathbf{q}},$
which possess the property \cite{Feynman}
\begin{equation}
Q_{\mathbf{q}}^{\ast}=Q_{-\mathbf{q}}.
\end{equation}
The full set of the electron and phonon coordinates are denoted by
$\mathbf{\bar{x}\equiv}\left\{  \mathbf{x}_{j,\sigma}\right\}  $ and $\bar
{Q}\equiv\left\{  Q_{\mathbf{q}}\right\}  .$

Throughout the sections \ref{sec:Freeenergy} and \ref{sec:groundstate}, the
Euclidean time variable $\tau=it$ is used, where $t$ is the real time
variable. In this representation the Lagrangian of the system is
\begin{equation}
L\left(  \mathbf{\dot{\bar{x}}},\dot{\bar{Q}};\mathbf{\bar{x}},\bar{Q}\right)
=L_{e}\left(  \mathbf{\dot{\bar{x}}},\mathbf{\bar{x}}\right)  -V_{C}\left(
\mathbf{\bar{x}}\right)  +L_{ph}\left(  \dot{\bar{Q}},\bar{Q}\right)
+L_{e-ph}\left(  \mathbf{\bar{x}},\bar{Q}\right)  ,\label{L}%
\end{equation}
where $L_{e}\left(  \mathbf{\dot{\bar{x}}},\mathbf{\bar{x}}\right)  $ is the
Lagrangian of an electron with band mass $m_{b}$ in a quantum dot:
\begin{equation}
L_{e}\left(  \mathbf{\dot{\bar{x}}},\mathbf{\bar{x}}\right)  =-\sum
_{\sigma=\pm1/2}\sum_{j=1}^{N_{\sigma}}\left(  \frac{m_{b}}{2}\mathbf{\dot{x}%
}_{j,\sigma}^{2}+\frac{m_{b}}{2}\Omega_{0}^{2}\mathbf{x}_{j,\sigma}%
^{2}\right)  ,\qquad\mathbf{\dot{x}\equiv}\frac{d\mathbf{x}}{d\tau},\label{Le}%
\end{equation}
$V_{C}\left(  \mathbf{\bar{x}}\right)  $ is the potential energy of the
electron-electron Coulomb repulsion in the medium with the high-frequency
dielectric constant $\varepsilon_{\infty}$:
\begin{equation}
V_{C}\left(  \mathbf{\bar{x}}\right)  =\underset{\left(  j,\sigma\right)
\neq\left(  l,\sigma^{\prime}\right)  }{\sum_{\sigma,\sigma^{\prime}=\pm
1/2}\sum_{j=1}^{N_{\sigma}}\sum_{l=1}^{N_{\sigma^{\prime}}}\frac{e^{2}%
}{2\varepsilon_{\infty}}}\frac{1}{\left|  \mathbf{x}_{j,\sigma}-\mathbf{x}%
_{l,\sigma^{\prime}}\right|  },\label{Vc}%
\end{equation}
$L_{ph}\left(  \dot{\bar{Q}},\dot{\bar{Q}}^{\ast};\bar{Q},\bar{Q}^{\ast
}\right)  $ is the Lagrangian of free phonons:
\begin{equation}
L_{ph}\left(  \dot{\bar{Q}},\bar{Q}\right)  =-\frac{1}{2}\sum_{\mathbf{q}%
}(\dot{Q}_{\mathbf{q}}^{\ast}\dot{Q}_{\mathbf{q}}+\omega_{\mathbf{q}}%
^{2}Q_{\mathbf{q}}^{\ast}Q_{\mathbf{q}}),\qquad\dot{Q}\mathbf{\equiv}%
\frac{dQ}{d\tau}.
\end{equation}
Further, $L_{e-ph}\left(  \mathbf{\bar{x}},\bar{Q},\bar{Q}^{\ast}\right)  $ is
the Lagrangian of the electron-phonon interaction:
\begin{equation}
L_{e-ph}\left(  \mathbf{\bar{x}},\bar{Q}\right)  =-\sum_{\mathbf{q}}\left(
\frac{2\omega_{\mathbf{q}}}{\hbar}\right)  ^{1/2}V_{\mathbf{q}}Q_{-\mathbf{q}%
}\rho_{\mathbf{q}},\label{Leph}%
\end{equation}
where $\rho_{\mathbf{q}}$ is the Fourier transform of the electron density
operator:%
\begin{equation}
\rho_{\mathbf{q}}=\sum_{\sigma=\pm1/2}\sum_{j=1}^{N_{\sigma}}e^{i\mathbf{q}%
\cdot\mathbf{x}_{j,\sigma}}.
\end{equation}
$V_{\mathbf{q}}$ is the amplitude of the electron-phonon interaction. In this
paper, we only consider electrons interacting with the long-wavelength
longitudinal optical (LO) phonons with a dispersionless frequency
$\omega_{\mathbf{q}}=\omega_{\mathrm{LO}}$, for which the amplitude
$V_{\mathbf{q}}$ is \cite{Green}%
\begin{equation}
V_{\mathbf{q}}=\frac{\hbar\omega_{\mathrm{LO}}}{q}\left(  \frac{2\sqrt{2}%
\pi\alpha}{V}\right)  ^{1/2}\left(  \frac{\hbar}{m_{b}\omega_{\mathrm{LO}}%
}\right)  ^{1/4},
\end{equation}
where $\alpha$ is the electron-phonon coupling constant and $V$ is the volume
of the crystal.

We treat a \emph{canonical} ensemble, where the numbers $N_{\sigma}$ are
fixed. The partition function $Z\left(  \left\{  N_{\sigma}\right\}
,\beta\right)  $ of the system can be expressed as a path integral over the
electron and phonon coordinates:
\begin{equation}
Z\left(  \left\{  N_{\sigma}\right\}  ,\beta\right)  =\sum_{P}\frac{\left(
-1\right)  ^{\mathbf{\xi}_{P}}}{N_{1/2}!N_{-1/2}!}\int d\mathbf{\bar{x}}%
\int_{\mathbf{\bar{x}}}^{P\mathbf{\bar{x}}}D\mathbf{\bar{x}}\left(
\tau\right)  \int d\bar{Q}\int_{\bar{Q}}^{\bar{Q}}D\bar{Q}\left(  \tau\right)
e^{-S\left[  \mathbf{\bar{x}}\left(  \tau\right)  ,\bar{Q}\left(  \tau\right)
\right]  },\label{Z}%
\end{equation}
where $S\left[  \mathbf{\bar{x}}\left(  \tau\right)  ,\bar{Q}\left(
\tau\right)  \right]  $ is the ``action'' functional:
\begin{equation}
S\left[  \mathbf{\bar{x}}\left(  \tau\right)  ,\bar{Q}\left(  \tau\right)
\right]  =-\frac{1}{\hbar}\int_{0}^{\hbar\beta}L\left(  \mathbf{\dot{\bar{x}}%
},\dot{\bar{Q}};\mathbf{\bar{x}},\bar{Q}\right)  d\tau.\label{S}%
\end{equation}
The parameter $\beta\equiv1/\left(  k_{B}T\right)  $ is inversely proportional
to the temperature $T$. In order to take the Fermi-Dirac statistics into
account, the integral over the electron paths $\left\{  \mathbf{\bar{x}%
}\left(  \tau\right)  \right\}  $ in Eq. (\ref{Z}) contains a sum over all
permutations $P$ of the electrons with the same spin projection, and
$\mathbf{\xi}_{P}$ denotes the parity of a permutation $P$.

The action functional (\ref{S}) is quadratic in the phonon coordinates
$\bar{Q}.$ Therefore, the path integral over the phonon variables in $Z\left(
\left\{  N_{\sigma}\right\}  ,\beta\right)  $ can be calculated analytically
\cite{Feynman}. As a result, the partition function of the electron-phonon
system (\ref{Z}) factorizes into a product%

\begin{equation}
Z\left(  \left\{  N_{\sigma}\right\}  ,\beta\right)  =Z_{p}\left(  \left\{
N_{\sigma}\right\}  ,\beta\right)  \prod_{\mathbf{q}}\frac{1}{2\sinh\left(
\beta\hbar\omega_{\mathrm{LO}}/2\right)  }\label{Factorized}%
\end{equation}
of the partition function of free phonons with a partition function
$Z_{p}\left(  \left\{  N_{\sigma}\right\}  ,\beta\right)  $ of interacting
polarons, which is a path integral over the electron coordinates only:%
\begin{equation}
Z_{p}\left(  \left\{  N_{\sigma}\right\}  ,\beta\right)  =\sum_{P}%
\frac{\left(  -1\right)  ^{\mathbf{\xi}_{P}}}{N_{1/2}!N_{-1/2}!}\int
d\mathbf{\bar{x}}\int_{\mathbf{\bar{x}}}^{P\mathbf{\bar{x}}}D\mathbf{\bar{x}%
}\left(  \tau\right)  e^{-S_{p}\left[  \mathbf{\bar{x}}\left(  \tau\right)
\right]  }.\label{Zp}%
\end{equation}
The functional
\begin{align}
S_{p}\left[  \mathbf{\bar{x}}\left(  \tau\right)  \right]   & =-\frac{1}%
{\hbar}\int_{0}^{\hbar\beta}\left[  L_{e}\left(  \mathbf{\dot{\bar{x}}}\left(
\tau\right)  ,\mathbf{\bar{x}}\left(  \tau\right)  \right)  +V_{C}\left(
\mathbf{\bar{x}}\left(  \tau\right)  \right)  \right]  d\tau\nonumber\\
& -\sum_{\mathbf{q}}\frac{\left|  V_{\mathbf{q}}\right|  ^{2}}{2\hbar^{2}}%
\int\limits_{0}^{\hbar\beta}d\tau\int\limits_{0}^{\hbar\beta}d\tau^{\prime
}\frac{\cosh\left[  \omega_{\mathrm{LO}}\left(  \left|  \tau-\tau^{\prime
}\right|  -\hbar\beta/2\right)  \right]  }{\sinh\left(  \beta\hbar
\omega_{\mathrm{LO}}/2\right)  }\rho_{\mathbf{q}}\left(  \tau\right)
\rho_{-\mathbf{q}}\left(  \tau^{\prime}\right) \label{Sp}%
\end{align}
results from the elimination of the phonon coordinates and contains the
``influence phase'' of the phonons (the last term in the right-hand side). It
describes the phonon-induced retarded interaction between the electrons,
including the retarded self-interaction of each electron. The free energy of a
system of interacting polarons $F_{p}\left(  \left\{  N_{\sigma}\right\}
,\beta\right)  $ is related to their partition function (\ref{Zp}) by the
equation:
\begin{equation}
F_{p}\left(  \left\{  N_{\sigma}\right\}  ,\beta\right)  =-\frac{1}{\beta}\ln
Z_{p}\left(  \left\{  N_{\sigma}\right\}  ,\beta\right)  .\label{Fp}%
\end{equation}

At present no method is known to calculate the non-gaussian path integral
(\ref{Zp}) analytically. For \emph{distinguishable} particles, the
Jensen-Feynman variational principle \cite{Feynman} provides a convenient
approximation technique. It yields a lower bound to the partition function,
and hence an upper bound to the free energy.

The formulation of a variational principle for the free energy for a system of
\emph{identical} \emph{particles }is a\emph{\ }non-trivial problem. However,
it can be shown \cite{PRE96} that the path-integral approach to the many-body
problem for a fixed number of identical particles can be formulated as a
Feynman-Kac functional on a state space for $N$ indistinguishable particles,
by imposing an ordering on the configuration space and by the introduction of
a set of boundary conditions at the boundaries of this state space. The path
integral (in the imaginary-time variable) for identical particles was shown to
be \emph{positive} within this state space. This implies that a many-body
extension of the Jensen-Feynman inequality was found, which can be used for
interacting identical particles (Ref.\cite{PRE96}, p. 4476). A more detailed
analysis of this variational principle for both local and retarded
interactions can be found in Ref. \cite{Note}. It is required that the
potentials are symmetric with respect to all permutations of the particle
positions, and that both the exact propagator and the model propagator are
antisymmetric (for fermions) with respect to permutations of any two electrons
at any moment in time. This means that these propagators have to be defined on
the same configuration space. Keeping in mind these constraints, the
variational inequality for identical particles takes the same form as the
Jensen-Feynman variational principle:
\begin{equation}
F_{p}\leqslant F_{0}+\frac{1}{\beta}\left\langle S_{p}-S_{0}\right\rangle
_{S_{0}},\label{JF}%
\end{equation}
where $S_{0}$ is a model action with corresponding free energy $F_{0}$. The
angular brackets mean a weighted average over the paths%
\begin{equation}
\left\langle \left(  \bullet\right)  \right\rangle _{S_{0}}=\frac{\sum
_{P}\frac{\left(  -1\right)  ^{\mathbf{\xi}_{P}}}{N_{1/2}!N_{-1/2}!}\int
d\mathbf{\bar{x}}\int_{\mathbf{\bar{x}}}^{P\mathbf{\bar{x}}}D\mathbf{\bar{x}%
}\left(  \tau\right)  \left(  \bullet\right)  e^{-S_{0}\left[  \mathbf{\bar
{x}}\left(  \tau\right)  \right]  }}{\sum_{P}\frac{\left(  -1\right)
^{\mathbf{\xi}_{P}}}{N_{1/2}!N_{-1/2}!}\int d\mathbf{\bar{x}}\int
_{\mathbf{\bar{x}}}^{P\mathbf{\bar{x}}}D\mathbf{\bar{x}}\left(  \tau\right)
e^{-S_{0}\left[  \mathbf{\bar{x}}\left(  \tau\right)  \right]  }}.\label{Aver}%
\end{equation}

\subsection{Model system}

We consider a model system consisting of $N$ electrons with coordinates
$\mathbf{\bar{x}\equiv}\left\{  \mathbf{x}_{j,\sigma}\right\}  $ and $N_{f}$
fictitious particles with coordinates $\mathbf{\bar{y}\equiv}\left\{
\mathbf{y}_{j}\right\}  $ in a harmonic confinement potential with elastic
interparticle interactions as studied in Ref. \cite{SSC114-305}. The
Lagrangian of this model system takes the form%
\begin{align}
L_{M}\left(  \mathbf{\dot{\bar{x}}},\mathbf{\dot{\bar{y}}};\mathbf{\bar{x}%
},\mathbf{\bar{y}}\right)   & =-\frac{m_{b}}{2}\sum_{\sigma}\sum
_{j=1}^{N_{\sigma}}\left(  \mathbf{\dot{x}}_{j,\sigma}^{2}+\Omega
^{2}\mathbf{x}_{j,\sigma}^{2}\right)  +\frac{m_{b}\omega^{2}}{4}\sum
_{\sigma,\sigma^{\prime}}\sum_{j=1}^{N_{\sigma}}\sum_{l=1}^{N_{\sigma^{\prime
}}}\left(  \mathbf{x}_{j,\sigma}-\mathbf{x}_{l,\sigma^{\prime}}\right)
^{2}\nonumber\\
& -\frac{m_{f}}{2}\sum_{j=1}^{N_{f}}\left(  \mathbf{\dot{y}}_{j}^{2}%
+\Omega_{f}^{2}\mathbf{y}_{j}^{2}\right)  -\frac{k}{2}\sum_{\sigma}\sum
_{j=1}^{N_{\sigma}}\sum_{l=1}^{N_{f}}\left(  \mathbf{x}_{j,\sigma}%
-\mathbf{y}_{l}\right)  ^{2}.\label{LM}%
\end{align}
The frequencies $\Omega,$ $\omega,$ $\Omega_{f},$ the mass of a fictitious
particle $m_{f},$ and the force constant $k$ are variational parameters.
Clearly, this Lagrangian is symmetric with respect to electron permutations.
Performing the path integral over the coordinates of the fictitious particles
\cite{Feynman}, the partition function $Z_{0}\left(  \left\{  N_{\sigma
}\right\}  ,\beta\right)  $ of the model system of interacting polarons
becomes a path integral over the electron coordinates:%
\begin{equation}
Z_{0}\left(  \left\{  N_{\sigma}\right\}  ,\beta\right)  =\sum_{P}%
\frac{\left(  -1\right)  ^{\mathbf{\xi}_{P}}}{N_{1/2}!N_{-1/2}!}\int
d\mathbf{\bar{x}}\int_{\mathbf{\bar{x}}}^{P\mathbf{\bar{x}}}D\mathbf{\bar{x}%
}\left(  \tau\right)  e^{-S_{0}\left[  \mathbf{\bar{x}}\left(  \tau\right)
\right]  },\label{Z0}%
\end{equation}
with the action functional $S_{0}\left[  \mathbf{\bar{x}}\left(  \tau\right)
\right]  $ given by%
\begin{align}
S_{0}\left[  \mathbf{\bar{x}}\left(  \tau\right)  \right]   & =\frac{1}{\hbar
}\int_{0}^{\hbar\beta}\sum_{\sigma}\sum_{j=1}^{N_{\sigma}}\frac{m_{b}}%
{2}\left[  \mathbf{\dot{x}}_{j,\sigma}^{2}\left(  \tau\right)  +\Omega
^{2}\mathbf{x}_{j,\sigma}^{2}\left(  \tau\right)  \right]  d\tau\nonumber\\
& -\frac{1}{\hbar}\int_{0}^{\hbar\beta}\sum_{\sigma,\sigma^{\prime}}\sum
_{j=1}^{N_{\sigma}}\sum_{l=1}^{N_{\sigma^{\prime}}}\frac{m_{b}\omega^{2}}%
{4}\left[  \mathbf{x}_{j,\sigma}\left(  \tau\right)  -\mathbf{x}%
_{l,\sigma^{\prime}}\left(  \tau\right)  \right]  ^{2}d\tau\nonumber\\
& -\frac{k^{2}N^{2}N_{f}}{4m_{f}\hbar\Omega_{f}}\int\limits_{0}^{\hbar\beta
}d\tau\int\limits_{0}^{\hbar\beta}d\tau^{\prime}\frac{\cosh\left[  \Omega
_{f}\left(  \left|  \tau-\tau^{\prime}\right|  -\hbar\beta/2\right)  \right]
}{\sinh\left(  \beta\hbar\Omega_{f}/2\right)  }\mathbf{X}\left(  \tau\right)
\cdot\mathbf{X}\left(  \tau^{\prime}\right)  ,\label{S0}%
\end{align}
where $\mathbf{X}$ is the center-of-mass coordinate of the electrons,%
\begin{equation}
\mathbf{X=}\frac{1}{N}\sum_{\sigma}\sum_{j=1}^{N_{\sigma}}\mathbf{x}%
_{j,\sigma}.\label{X}%
\end{equation}
The details of the analytical calculation of the model partition function
(\ref{Z0}) are described in Appendix A.

After substituting the model action functional (\ref{S0}) into the right-hand
side of the variational inequality (\ref{JF}), we obtain an upper bound to the
free energy $F_{var},$%
\begin{align}
& F_{var}\left(  \left\{  N_{\sigma}\right\}  ,\beta\right) \nonumber\\
& =F_{0}\left(  \left\{  N_{\sigma}\right\}  ,\beta\right)  +\frac{m_{b}}%
{2}\left(  \Omega_{0}^{2}-\Omega^{2}+N\omega^{2}\right)  \left\langle
\sum_{j=1}^{N}\mathbf{x}_{j}^{2}\left(  0\right)  \right\rangle _{S_{0}%
}\nonumber\\
& -\frac{m_{b}\omega^{2}N^{2}}{2}\left\langle \mathbf{X}^{2}\left(  0\right)
\right\rangle _{S_{0}}+\sum_{\mathbf{q}\neq0}\frac{2\pi e^{2}}{V\varepsilon
_{\infty}q^{2}}\left[  g\left(  \mathbf{q},0|\left\{  N_{\sigma}\right\}
,\beta\right)  -N\right] \nonumber\\
& +\frac{k^{2}N^{2}N_{f}}{4m_{f}\beta\hbar\Omega_{f}}\int\limits_{0}%
^{\hbar\beta}d\tau\int\limits_{0}^{\hbar\beta}d\tau^{\prime}\frac{\cosh\left[
\Omega_{f}\left(  \left|  \tau-\tau^{\prime}\right|  -\hbar\beta/2\right)
\right]  }{\sinh\left(  \beta\hbar\Omega_{f}/2\right)  }\left\langle
\mathbf{X}\left(  \tau\right)  \cdot\mathbf{X}\left(  \tau^{\prime}\right)
\right\rangle _{S_{0}}\nonumber\\
& -\sum_{\mathbf{q}}\frac{\left|  V_{\mathbf{q}}\right|  ^{2}}{2\hbar^{2}%
\beta}\int\limits_{0}^{\hbar\beta}d\tau\int\limits_{0}^{\hbar\beta}%
d\tau^{\prime}\frac{\cosh\left[  \omega_{\mathrm{LO}}\left(  \left|  \tau
-\tau^{\prime}\right|  -\hbar\beta/2\right)  \right]  }{\sinh\left(
\beta\hbar\omega_{\mathrm{LO}}/2\right)  }g\left(  \mathbf{q},\tau
-\tau^{\prime}|\left\{  N_{\sigma}\right\}  ,\beta\right)  .\label{varfun}%
\end{align}
Here, $g\left(  \mathbf{q},\tau-\tau^{\prime}|\left\{  N_{\sigma}\right\}
,\beta\right)  $ is the two-point correlation function for the electron
density operators:%
\begin{equation}
g\left(  \mathbf{q},\tau|\left\{  N_{\sigma}\right\}  ,\beta\right)
=\left\langle \rho_{\mathbf{q}}\left(  \tau\right)  \rho_{-\mathbf{q}}\left(
0\right)  \right\rangle _{S_{0}}.\label{TPCF}%
\end{equation}
Both the free energy and the correlation functions of the model system can be
calculated analytically using the generating function technique \cite{PRE97}.
In the zero-temperature limit $\left(  \beta\rightarrow\infty\right)  ,$ the
variational free energy (\ref{varfun}) becomes an upper bound $E_{var}%
^{0}\left(  \left\{  N_{\sigma}\right\}  \right)  $ to the ground-state energy
$E^{0}$ of the system of interacting polarons. The details of the calculation
of the correlation functions are given in the Appendix B.

\section{Ground-state energy and addition energy of interacting polarons}

\label{sec:groundstate}

For the numerical calculations, we use effective atomic units, where $\hbar, $
the electron band mass $m_{b}$ and $e/\sqrt{\varepsilon_{\infty}}$ have the
numerical value of 1. This means that the unit of length is the effective Bohr
radius $a_{B}^{\ast}=\hbar^{2}\varepsilon_{\infty}/\left(  m_{b}e^{2}\right)
$, while the unit of energy is the effective Hartree\cite{PRB62-8108}
$H^{\ast}=m_{b}e^{4}/\left(  \hbar^{2}\varepsilon_{\infty}^{2}\right)  $.
These units allow to present results for quantum dots with and without the
electron-phonon interaction on the same scale. Therefore, for confined
polarons they are more convenient than the usual polaron units, where the unit
of length is $a_{p}\equiv\left[  \hbar/\left(  m_{b}\omega_{\mathrm{LO}%
}\right)  \right]  ^{1/2},$ and the energy is measured in units of the
LO-phonon energy $\hbar\omega_{\mathrm{LO}}.$ In terms of the dimensionless
parameters $\alpha$ and $\eta\equiv\varepsilon_{\infty}/\varepsilon_{0},$where
$\varepsilon_{0}$ is the static dielectric constant, the following relations
exist between both systems of units:%
\begin{equation}
\frac{a_{p}}{a_{B}^{\ast}}=\frac{\sqrt{2}\alpha}{1-\eta},\quad\frac{H^{\ast}%
}{\hbar\omega_{\mathrm{LO}}}=\left(  \frac{a_{p}}{a_{B}^{\ast}}\right)
^{2}=\frac{2\alpha^{2}}{\left(  1-\eta\right)  ^{2}}.\label{rrr}%
\end{equation}

In Fig. 1, the total spin $S$\ of a system of interaction polarons in their
ground state is plotted as a function of the number of electrons in a 3D
quantum dot for different values of the confinement frequency $\Omega_{0}$, of
the electron-phonon coupling constant $\alpha$ and of the parameter $\eta
.$\ As distinct from few-electron systems without the electron-phonon
interaction, three types of spin polarization are possible for the ground
state, which should be distinguishable from each other using, e.g., capacity measurements.

(i) Except for the strong-coupling case and for the low-density case, the
filling in the ground state is as follows: in an open shell, with
less-than-half filling, each new electron is added with one and the same spin,
so that the total spin (in the shell under consideration) is maximal in
accordance with Hund's rule \cite{Messiah}. As soon as half-filling is
achieved with electrons possessing a certain spin, each new electron is added
with the spin opposite to that in the group of electrons providing the
aforementioned half-filling. When the number of electrons corresponds to the
number of states in the shell under consideration, the shell becomes closed,
and the total spin is zero. This mode of filling is referred to as Hund's rule
for a quantum dot. Hund's rule means, that the electrons in a partly filled
upper shell build up a minimal possible number of pairs in order to minimize
the electron-electron repulsion. For a quantum dot with $\hbar\Omega_{0}=0.5$
$H^{\ast}$ at $\alpha=0$ and at $\alpha=0.5$, the shell filling always obeys
Hund's rule, as shown in Fig. 1(\emph{a}).

(ii) With decreasing confinement frequency $\Omega_{0}$ at a given number of
electrons, the electron density lowers. For densities smaller than a certain
value, it can happen that the ground state is a state with a maximal total (in
all shells) spin. In this state, the electrons are filling consecutively all
the single-electron states with one and the same spin and are referred to as
spin-polarized. The examples are the states at $\alpha=0$ for $N=4$ and $N=10$
in Fig. 1(\emph{b}). A spin-polarized ground state precedes the formation of a
Wigner crystal when further lowering the density \cite{PRB62-8108,AJP49-161}.

(iii) In the strong-coupling case ( $\alpha\gg1$ and $\eta\ll1$), it can
happen that the ground state is a state with a minimal total spin (0 for even
number of electrons and
${\frac12}$
for odd number of electrons). This is the case when --due to the
phonon-mediated electron-electron attraction-- pairing of electrons with
opposite spins occurs, analogous to a singlet bipolaron ground state in bulk.
The examples are the states at $\alpha=5$ and $\eta=0.1$ for $N$ in the range
from 4 to 6 in Fig. 1(\emph{a}) and for $N$ in the range from 4 to 10 in Fig.
1(\emph{b}). This trend to minimize the total spin is a consequence of the
electron-phonon interaction, presumably due to the fact that the
phonon-mediated electron-electron attraction overcomes the Coulomb repulsion.
With an increasing number of electrons, at a certain value of $N$, such states
with a minimal total spin cease to form the ground state, and the shell
filling abruptly returns to that prescribed by Hund's rule [see a jump in the
spin at $\alpha=5$ when $N$ changes from 10 to 11 in Fig. 1(\emph{b})]. This
jump is analogous to a transition from states with paired electrons (like
superconducting states) to another type of states with unpaired electrons
(like normal states).

The addition energy $\Delta\left(  N\right)  $, which is the variation of the
chemical potential when putting an extra electron into a quantum dot, is
defined as \cite{PRB57-9035,PRB59-4604}%
\begin{equation}
\Delta\left(  N\right)  =E^{0}\left(  N+1\right)  -2E^{0}\left(  N\right)
+E^{0}\left(  N-1\right)  .\label{Add}%
\end{equation}
Fig. 2 presents the addition energy in a 3D quantum dot as a function of the
number of electrons. The structure of $\Delta\left(  N\right)  $ clearly
manifests the shell structure of a quantum dot. The most pronounced peaks in
the addition energy occur for closed-shell systems with $N=2,8,20$. The peaks
in $\Delta\left(  N\right)  $ at $N=5$ and $N=14$ obtained within the present
approach for relatively weak electron-phonon coupling correspond to the
systems with the half-filled upper shell (see Fig. 2(\emph{a}) for $\alpha=0$
and $\alpha=0.5$). In these cases the total spin for the upper shell takes its
maximal possible value, in accordance with Hund's rule. At sufficiently large
values of $\alpha,$ the electron-phonon interaction substantially modifies the
addition energy. In the strong-coupling case, the peaks corresponding to
half-filled shells become less pronounced, while those corresponding to
closed-shell systems become more prominent as compared to the weak-coupling case.

To the best of our knowledge, the addition energy for parabolic quantum dots
was obtained using the density functional theory (DFT) (see, e.g.,
Refs.\cite{PRB63-045317,PRB57-9035,PRB59-4604}) only without the
electron-phonon interaction. Our results for the addition energy for a 3D
quantum dot as a function of $N$ in the particular case $\alpha=0$ are very
close for $N\leqslant12$ to those calculated within the DFT
\cite{PRB63-045317} with an optimized effective potential and a
self-interaction correction \cite{PRA55-3406}.

The panels \emph{a} and \emph{b} in Fig. 3 represent, respectively, the total
spin and the addition energy for interacting polarons in a 2D parabolic GaAs
quantum dot with the confinement parameter $\hbar\Omega_{0}=0.5\,H^{\ast
}\approx7.67$ meV. The pronounced peaks in $\Delta\left(  N\right)  $ at
$N=2,6,12,20,\ldots$ correspond to the closed-shell systems, for which the
total spin equals zero. In accordance with Hund's rule, the upper shell is
filled in such a way that the total spin of electrons in this shell takes the
maximal possible value. Therefore for the half-filled upper shell (at
$N=4,9,16,\ldots$) maxima of the total spin occur as a function of $N.$ At
these electron numbers, the addition energy manifests peaks, which are less
pronounced than those corresponding to closed shells.

The inset to Fig. 3(\emph{b}) shows the experimental data for the addition
energy in a cylindric GaAs quantum dot \cite{Tarucha1996}. As seen from Fig.
3, the peak positions for the addition energy of interacting polarons in a 2D
parabolic quantum dot agree well with the experimental results for the
addition energies of cylindrical quantum dots. The height of the calculated
peaks of the addition energy falls down as the shell number increases, which
is qualitatively consistent with the experimentally observed behavior
\cite{Tarucha1996}. The peaks in $\Delta\left(  N\right)  $ corresponding to
the half-filled shells are weaker than those for the closed shells both in the
experiment\cite{Tarucha1996} and in our theory.

\section{Optical conductivity}

\label{sec:optconductivity}

For a system of interacting polarons in a parabolic confinement potential, we
calculate the real part of the optical conductivity within the memory-function
approach. For a single polaron at arbitrary coupling strength it was developed
in Refs. \cite{DSG,Green}. For a polaron gas in the weak-coupling limit, this
technique was applied in Ref. \cite{Wu1986}.

In the present paper we extend the memory-function approach to a system of
arbitrary-coupling interacting polarons. Since the optical conductivity
relates the current $\mathbf{J}\left(  t\right)  $ per electron to a
time-dependent uniform electric field $\mathbf{E}\left(  t\right)  $ in the
framework of linear response theory, we have to return to the \emph{real-time}
representation in the path integrals. The Fourier components of the electric
field are denoted by $\mathbf{E}_{\omega}:$
\begin{equation}
\mathbf{E}\left(  t\right)  =\frac{1}{2\pi}\int_{-\infty}^{\infty}%
\mathbf{E}_{\omega}e^{-i\omega t}d\omega,\label{Fourier}%
\end{equation}
and the similar denotations are used for other time-dependent quantities. The
electric current per electron $\mathbf{J}\left(  t\right)  $ is related to the
mean electron coordinate response $\mathbf{R}\left(  t\right)  $ by
\begin{equation}
\mathbf{J}\left(  t\right)  =-e\frac{d\mathbf{R}\left(  t\right)  }%
{dt},\label{J}%
\end{equation}
and hence%
\begin{equation}
\mathbf{J}_{\omega}=ie\omega\mathbf{R}_{\omega}.\label{J1}%
\end{equation}
Within the linear-response theory, both the electric current and the
coordinate response are proportional to $\mathbf{E}_{\omega}$:%
\begin{equation}
\mathbf{J}_{\omega}=\sigma\left(  \omega\right)  \mathbf{E}_{\omega}%
,\quad\mathbf{R}_{\omega}=\frac{\sigma\left(  \omega\right)  }{ie\omega
}\mathbf{E}_{\omega},\label{J2}%
\end{equation}
where $\sigma\left(  \omega\right)  $ is the conductivity per electron.
Because we treat an isotropic electron-phonon system, $\sigma\left(
\omega\right)  $ is a scalar function. It is determined from the time
evolution of the center-of-mass coordinate:
\begin{equation}
\mathbf{R}\left(  t\right)  \equiv\frac{1}{N}\left\langle \left\langle
\sum_{j=1}^{N}\mathbf{x}_{j}\left(  t\right)  \right\rangle \right\rangle
_{S}.\label{R}%
\end{equation}
The symbol $\left\langle \left\langle \left(  \bullet\right)  \right\rangle
\right\rangle _{S}$ denotes an average in the \emph{real-time} representation
for a system with action functional $S$:%
\begin{equation}
\left\langle \left\langle \left(  \bullet\right)  \right\rangle \right\rangle
_{S}\equiv\int d\mathbf{\bar{x}}\int d\mathbf{\bar{x}}_{0}\int d\mathbf{\bar
{x}}_{0}^{\prime}\int\limits_{\mathbf{\bar{x}}_{0}}^{\mathbf{\bar{x}}%
}D\mathbf{\bar{x}}\left(  t\right)  \int\limits_{\mathbf{\bar{x}}_{0}^{\prime
}}^{\mathbf{\bar{x}}}D\mathbf{\bar{x}}^{\prime}\left(  t\right)
e^{\frac{i}{\hbar}S\left[  \mathbf{\bar{x}}\left(  t\right)  ,\mathbf{\bar{x}%
}^{\prime}\left(  t\right)  \right]  }\left(  \bullet\right)  \left.
\left\langle \mathbf{\bar{x}}_{0}\left|  \hat{\rho}\left(  t_{0}\right)
\right|  \mathbf{\bar{x}}_{0}^{^{\prime}}\right\rangle \right|  _{t_{0}%
\rightarrow-\infty},\label{Average}%
\end{equation}
where $\left\langle \mathbf{\bar{x}}_{0}\left|  \hat{\rho}\left(
t_{0}\right)  \right|  \mathbf{\bar{x}}_{0}^{^{\prime}}\right\rangle $ is the
density matrix before the onset of the electric field in the infinite past
$\left(  t_{0}\rightarrow-\infty\right)  $. The corresponding action
functional is \cite{TF,DF96}%
\begin{equation}
S\left[  \mathbf{\bar{x}}\left(  t\right)  ,\mathbf{\bar{x}}^{\prime}\left(
t\right)  \right]  =\int\limits_{-\infty}^{t}\left[  L_{e}\left(
\mathbf{\dot{\bar{x}}}\left(  t\right)  ,\mathbf{\bar{x}}\left(  t\right)
,t\right)  -L_{e}\left(  \mathbf{\dot{\bar{x}}}^{\prime}\left(  t\right)
,\mathbf{\bar{x}}^{\prime}\left(  t\right)  ,t\right)  \right]  dt^{\prime
}-i\hbar\Phi\left[  \mathbf{\bar{x}}\left(  t\right)  ,\mathbf{\bar{x}%
}^{\prime}\left(  t\right)  \right]  ,\label{S1}%
\end{equation}
where $L_{e}\left(  \mathbf{\dot{\bar{x}}},\mathbf{\bar{x}},t\right)  $ is the
Lagrangian of $N$ interacting electrons in a time-dependent uniform electric
field $\mathbf{E}\left(  t\right)  $
\begin{equation}
L_{e}\left(  \mathbf{\dot{\bar{x}}},\mathbf{\bar{x}},t\right)  =\sum_{\sigma
}\sum_{j=1}^{N_{\sigma}}\left(  \frac{m_{b}\mathbf{\dot{x}}_{j,\sigma}^{2}}%
{2}-\frac{m_{b}\Omega_{0}^{2}\mathbf{x}_{j,\sigma}^{2}}{2}-e\mathbf{x}%
_{j,\sigma}\cdot\mathbf{E}\left(  t\right)  \right)  -\underset{\left(
j,\sigma\right)  \neq\left(  l,\sigma^{\prime}\right)  }{\sum_{\sigma
,\sigma^{\prime}}\sum_{j=1}^{N_{\sigma}}\sum_{l=1}^{N_{\sigma^{\prime}}}%
}\frac{e^{2}}{2\varepsilon_{\infty}\left|  \mathbf{x}_{j,\sigma}%
-\mathbf{x}_{l,\sigma^{\prime}}\right|  }.\label{L1}%
\end{equation}
The influence phase of the phonons (see, e.g., Ref. \cite{DF96})
\begin{align}
\Phi\left[  \mathbf{\bar{x}}\left(  s\right)  ,\mathbf{\bar{x}}^{\prime
}\left(  s\right)  \right]   & =-\sum_{\mathbf{q}}\frac{\left|  V_{\mathbf{q}%
}\right|  ^{2}}{\hbar^{2}}\int\limits_{-\infty}^{t}ds\int\limits_{-\infty}%
^{s}ds^{\prime}\left[  \rho_{\mathbf{q}}\left(  s\right)  -\rho_{\mathbf{q}%
}^{\prime}\left(  s\right)  \right] \nonumber\\
& \times\left[  T_{\omega_{\mathbf{q}}}^{\ast}\left(  s-s^{\prime}\right)
\rho_{\mathbf{q}}\left(  s^{\prime}\right)  -T_{\omega_{\mathbf{q}}}\left(
s-s^{\prime}\right)  \rho_{\mathbf{q}}^{\prime}\left(  s^{\prime}\right)
\right] \label{Phi}%
\end{align}
describes both a retarded interaction between different electrons and a
retarded self-interaction of each electron due to the elimination of the
phonon coordinates. This functional contains the free-phonon Green's
function:
\begin{equation}
T_{\omega}\left(  t\right)  =\frac{e^{i\omega t}}{1-e^{-\beta\hbar\omega}%
}+\frac{e^{-i\omega t}}{e^{\beta\hbar\omega}-1}.\label{PGF}%
\end{equation}
The equation of motion for $\mathbf{R}\left(  t\right)  $ can be derived by
analogy with that described in Ref. \cite{FH}:%
\begin{equation}
m_{b}\frac{d^{2}\mathbf{R}\left(  t\right)  }{dt^{2}}+m_{b}\Omega_{0}%
^{2}\mathbf{R}\left(  t\right)  +e\mathbf{E}\left(  t\right)  =\mathbf{F}%
_{ph}\left(  t\right)  ,\label{EqMotion}%
\end{equation}
where $\mathbf{F}_{ph}\left(  t\right)  $ is the average force due to the
electron-phonon interaction,
\begin{equation}
\mathbf{F}_{ph}\left(  t\right)  =-\operatorname{Re}\sum_{\mathbf{q}%
}\frac{2\left|  V_{\mathbf{q}}\right|  ^{2}\mathbf{q}}{N\hbar}\int
\limits_{-\infty}^{t}ds\;T_{\omega_{\mathrm{LO}}}^{\ast}\left(  t-s\right)
\left\langle \left\langle \rho_{\mathbf{q}}\left(  t\right)  \rho
_{-\mathbf{q}}\left(  s\right)  \right\rangle \right\rangle _{S}.\label{F}%
\end{equation}
The two-point correlation function $\langle\left\langle \rho_{\mathbf{q}%
}\left(  t\right)  \rho_{-\mathbf{q}}\left(  s\right)  \right\rangle
\rangle_{S}$ should be calculated from Eq.~(\ref{Average}) using the exact
action (\ref{S1}), but like for the free energy above, this path integral
cannot be calculated analytically. Instead, we perform an approximate
calculation, replacing the two-point correlation function in Eq. (\ref{F}) by
$\langle\left\langle \rho_{\mathbf{q}}\left(  t\right)  \rho_{-\mathbf{q}%
}\left(  s\right)  \right\rangle \rangle_{S_{0}},$ where $S_{0}\left[
\mathbf{\bar{x}}\left(  t\right)  ,\mathbf{\bar{x}}^{\prime}\left(  t\right)
\right]  $ is the action functional with the optimal values of the variational
parameters for the model system considered in the previous section in the
presence of the electric field $\mathbf{E}\left(  t\right)  $. The functional
$S_{0}\left[  \mathbf{\bar{x}}\left(  t\right)  ,\mathbf{\bar{x}}^{\prime
}\left(  t\right)  \right]  $ is quadratic and describes a system of coupled
harmonic oscillators in the uniform electric field $\mathbf{E}\left(
t\right)  $. This field enters the term $-e\mathbf{E}\left(  t\right)
\cdot\sum_{\sigma}\sum_{j=1}^{N_{\sigma}}\mathbf{x}_{j,\sigma}$ in the
Lagrangian, which only affects the center-of-mass coordinate. Hence, a shift
of variables to the frame of reference with the origin at the center of mass
\begin{equation}
\left\{
\begin{array}
[c]{l}%
\mathbf{x}_{n}\left(  t\right)  =\mathbf{\tilde{x}}_{n}\left(  t\right)
+\mathbf{R}\left(  t\right)  ,\\
\mathbf{x}_{n}^{\prime}\left(  t\right)  =\mathbf{\tilde{x}}_{n}^{\prime
}\left(  t\right)  +\mathbf{R}\left(  t\right)  ,
\end{array}
\right. \label{Tr1}%
\end{equation}
results in \cite{DF96}
\begin{equation}
\left\langle \left\langle \rho_{q}\left(  t\right)  \rho_{-q}\left(  s\right)
\right\rangle \right\rangle _{S_{0}}=\left.  \left\langle \left\langle
\rho_{q}\left(  t\right)  \rho_{-q}\left(  s\right)  \right\rangle
\right\rangle _{S_{0}}\right|  _{E=0}e^{i\mathbf{q}\cdot\left[  \mathbf{R}%
\left(  t\right)  -\mathbf{R}\left(  s\right)  \right]  }.\label{Tr2}%
\end{equation}
This result (\ref{Tr2}) is valid for any \emph{quadratic} model action $S_{0}. $

The applicability of the parabolic approximation for $N=1$ is confirmed by the
fact, that for the polaron ground-state energy, the results of the Feynman
approach \cite{Feynman55} are very close to the values obtained using other
reliable methods \cite{Pekar,Miyake,Adamowski,Mishchenko,DeFilippis}. Thus, a
self-induced polaronic potential, created by the polarization cloud around an
electron, is rather well described by a parabolic potential whose parameters
are determined by a variational method. For $N=2,$ the lowest known values of
the bipolaron ground-state energy are provided by the path-integral
variational method with parabolic potentials both in bulk \cite{VPD} and for
confined systems \cite{DW2001,EQD} for realistic values of $\alpha$. The
aforesaid approximation for the right-hand side of Eq. (\ref{F}) is a direct
generalization of the all-coupling approach \cite{DSG,DF96,Bipabs} to a
many-polaron system. For weak coupling, our variational method is at least of
the same accuracy as the perturbation theory, which results from our approach
at a special choice of the variational parameters. For strong coupling, an
interplay of the electron-phonon interaction and the Coulomb correlations
within a confinement potential can lead to the assemblage of polarons in
multi-polaron systems. As shown in Refs. \cite{DW2001,EQD} for a system with
$N=2$, the presence of a confinement potential strongly favors the bipolaron
formation. Our choice of the model variational system is reasonable because of
this trend, apparently occurring in a many-polaron system with arbitrary $N$
for a finite confinement strength.

The correlation function $\left.  \left\langle \rho_{\mathbf{q}}\left(
t\right)  \rho_{-\mathbf{q}}\left(  s\right)  \right\rangle _{S_{0}}\right|
_{\mathbf{E}=0}$ corresponds to the model system in the absence of an electric
field. For $t>s,$ this function is related to the imaginary-time correlation
function $g\left(  \mathbf{q},\tau|\left\{  N_{\sigma}\right\}  ,\beta\right)
,$ described in the previous section:%
\begin{equation}
\left.  \left\langle \left\langle \rho_{\mathbf{q}}\left(  t\right)
\rho_{-\mathbf{q}}\left(  s\right)  \right\rangle \right\rangle _{S_{0}%
}\right|  _{\mathbf{E}=0,t>s}=g\left(  \mathbf{q},i\left(  t-s\right)
|\left\{  N_{\sigma}\right\}  ,\beta\right)  .\label{Tr3}%
\end{equation}
Using the transformation (\ref{Tr1}) and the relation (\ref{Tr3}), one readily
obtains%
\begin{equation}
\mathbf{F}_{ph}\left(  t\right)  =-\operatorname{Re}\sum_{\mathbf{q}%
}\frac{2\left|  V_{\mathbf{q}}\right|  ^{2}\mathbf{q}}{N\hbar}\int
\limits_{-\infty}^{t}T_{\omega_{\mathrm{LO}}}^{\ast}\left(  t-s\right)
\mathrm{e}^{\mathrm{i}\mathbf{q}\cdot\left[  \mathbf{R}\left(  t\right)
-\mathbf{R}\left(  s\right)  \right]  }g\left(  \mathbf{q},i\left(
t-s\right)  |\left\{  N_{\sigma}\right\}  ,\beta\right)  ds.\label{F1}%
\end{equation}

Within the framework of the linear-response theory, the external electric
field $\mathbf{E}\left(  t\right)  $ is a small perturbation, so that
$\mathbf{R}\left(  t\right)  $ is a linear functional of $\left.  \left[
\mathbf{E}\left(  t^{\prime}\right)  \right]  \right|  _{t^{\prime}\leqslant
t}.$ Expanding the function $e^{i\mathbf{q}\cdot\left[  \mathbf{R}\left(
t\right)  -\mathbf{R}\left(  s\right)  \right]  }$ in the right-hand side of
Eq. (\ref{F1}) in powers of $\left[  \mathbf{R}\left(  t\right)
-\mathbf{R}\left(  s\right)  \right]  $ up to the first-order term, we obtain
the Fourier component $\mathbf{F}_{ph}\left(  \omega\right)  $ of the force
due to the electron-phonon interaction which is proportional to $\mathbf{R}%
_{\omega}.$ As a result, the optical conductivity can be expressed in terms of
the memory function $\chi\left(  \omega\right)  $ (cf. Refs. \cite{DSG,Green}%
),%
\begin{equation}
\operatorname{Re}\sigma\left(  \omega\right)  =-\frac{e^{2}}{m_{b}%
}\frac{\omega\operatorname{Im}\chi\left(  \omega\right)  }{\left[  \omega
^{2}-\Omega_{0}^{2}-\operatorname{Re}\chi\left(  \omega\right)  \right]
^{2}+\left[  \operatorname{Im}\chi\left(  \omega\right)  \right]  ^{2}%
},\label{Kw}%
\end{equation}
where $\chi\left(  \omega\right)  $ is given by
\begin{equation}
\chi\left(  \omega\right)  =\sum_{\mathbf{q}}\frac{2\left|  V_{\mathbf{q}%
}\right|  ^{2}q^{2}}{3N\hbar m_{b}}\int\limits_{0}^{\infty}dt\,\left(
e^{i\omega t}-1\right)  \operatorname{Im}\left[  T_{\omega_{\mathrm{LO}}%
}^{\ast}\left(  t\right)  g\left(  \mathbf{q},it|\left\{  N_{\sigma}\right\}
,\beta\right)  \right]  .\label{Hi}%
\end{equation}
It is worth noting that the optical conductivity (\ref{Kw}) differs from that
for a translationally invariant polaron system both by the explicit form of
$\chi\left(  \omega\right)  $ and by the presence of the term $\Omega_{0}^{2}$
in the denominator. For $\alpha\rightarrow0,$ the optical conductivity tends
to a $\delta$-like peak at $\omega=\Omega_{0},$%
\begin{equation}
\lim_{\alpha\rightarrow0}\operatorname{Re}\sigma\left(  \omega\right)
=\frac{\pi e^{2}}{2m_{b}}\delta\left(  \omega-\Omega_{0}\right)
.\label{Limit}%
\end{equation}
For a translationally invariant system $\Omega_{0}\rightarrow0$, and this
weak-coupling expression (\ref{Limit}) reproduces the ``central peak'' of the
polaron optical conductivity \cite{DLR77}.

In the zero-temperature limit, the memory function of Eq.~(\ref{Hi}) is
derived in the analytical form of Eq. (\ref{mf}) in Appendix B for 3D and 2D
interacting polarons.

\section{Results on the optical conductivity}

\label{sec:resultsconductivity}


Due to the confinement, the electron motion in a quantum dot is fully
quantized. Hence, when a photon is absorbed, the electron recoil can be
transferred only by discrete quanta. As a result, the optical conductivity
spectrum of a system of interacting polarons in a quantum dot is a series of
$\delta$-like peaks as distinct from the optical conductivity spectrum of a
bulk polaron \cite{DSG,Green}. These peaks are related to the internal polaron excitations.

Because $\operatorname{Im}\chi\left(  \omega\right)  =0$ for all frequencies
except for a discrete set of combinatorial frequencies (\ref{oklm}), the peaks
in the optical conductivity (\ref{Kw}) are positioned at the frequencies which
are given by the roots of the equation%
\begin{equation}
\omega^{2}-\Omega_{0}^{2}-\operatorname{Re}\chi\left(  \omega\right)
=0,\label{peaks}%
\end{equation}
which are denoted as $\left(  \tilde{\Omega}_{1},\tilde{\Omega}_{2}%
,\ldots\right)  $.

One of these roots is close to the variational parameter $\Omega_{2}$, which
is the eigenfrequency of the motion of a polaron as a whole. It satisfies the
inequality $\Omega_{2}<\Omega_{0}$ because the polaron effective mass is
larger than that of a bare electron. $\Omega_{2}$ is close to $\Omega_{0}\ $
in the weak-coupling case and decreases with increasing $\alpha.$ Hence, it
tends to zero in the limit $\Omega_{0}\rightarrow0.$ The peak in
$\operatorname{Re}\sigma\left(  \omega\right)  $ corresponding to this root
can be considered as the \emph{zero-phonon} line, which is an analogue of the
``central peak'' of the polaron optical conductivity \cite{DSG,Green}. The
peaks of $\mathop{\rm Re}  \sigma\left(  \omega\right) $ determined by the
other roots of Eq.~(\ref{peaks}) can be attributed to transitions into excited
states of the many-polaron system.

The changes of the shell filling schemes, which occur when varying the
confinement frequency, manifest themselves in the spectra of the optical
conductivity. In Fig.\thinspace4, optical conductivity spectra for $N=20$
polarons are presented for a quantum dot with the parameters of CdSe:
$\alpha=0.46,$ $\eta=0.656$ \cite{Kartheuser1972} and with different values of
the confinement energy $\hbar\Omega_{0}$. In this case, the spin-polarized
ground state changes to the ground state satisfying Hund's rule with
increasing $\hbar\Omega_{0}$ in the interval $0.0421H^{\ast}<\hbar\Omega
_{0}<0.0422H^{\ast}$.

In the inset to Fig.\thinspace4, the first frequency moment of the optical
conductivity
\begin{equation}
\left\langle \omega\right\rangle \equiv\frac{\int_{0}^{\infty}\omega
\operatorname{Re}\sigma\left(  \omega\right)  d\omega}{\int_{0}^{\infty
}\operatorname{Re}\sigma\left(  \omega\right)  d\omega},\label{Moment}%
\end{equation}
as a function of $\hbar\Omega_{0}$ shows a \emph{discontinuity}, at the value
of the confinement energy corresponding to the change of the shell filling
schemes from the spin-polarized ground state to the ground state obeying
Hund's rule. This discontinuity should be observable in optical measurements.

In Fig. 5, the first frequency moment (\ref{Moment}) is plotted as a function
of the number of electrons for a CdSe quantum dot with $\Omega_{0}%
=0.143\omega_{\mathrm{LO}}$ (corresponding to $\hbar\Omega_{0}\approx0.04$
$H^{\ast}$). The total spin of the system as a function of $N$ is shown in the
inset. As a general trend, $\left\langle \omega\right\rangle $ decreases with
increasing $N,$ with kinks corresponding to the ground-state transitions from
states obeying Hund's rule with $N=3,$ 9, and 18, into spin-polarized states
with $N=4,$ 10, and 19, respectively.

In Fig. 6, optical conductivity spectra are plotted for several values of the
confinement frequency for $N=10$ polarons in a quantum dot with $\alpha=2,$
$\eta=0.6$. These values of $\alpha$ and $\eta$ are typical for the
high-T$_{c}$ superconducting cuprates of the NCCO family \cite{TD1}. In the
``weak-confinement'' region ($\Omega_{0}=0.6\omega_{\mathrm{LO}}$ and
$\Omega_{0}=0.8\omega_{\mathrm{LO}}$) the zero-phonon peak is expressively
dominant over the other peaks.

When the confinement frequency parameter passes through the value $\Omega
_{0}=\omega_{\mathrm{LO}}$, the so-called ``confinement-phonon resonance''
\cite{SSC114-305} occurs. In this case, the peaks at $\tilde{\Omega}%
_{k},k=1,2,3$ have comparable oscillator strengths. The position
$\tilde{\Omega}_{2}$ of the second peak is substantially shifted from the LO
phonon frequency $\omega_{\mathrm{LO}}.$ Moreover, the intensities of the
phonon-assisted transitions increase as compared to the ``weak-confinement''
case. This resonance has a clear analogy with the magneto-phonon resonance
(see, e.g., Ref. \cite{PRL66-794}), as far as the energy levels of an electron
in a parabolic confinement are similar to the Landau levels of an electron in
a magnetic field.

With further increasing $\Omega_{0},$ when $\Omega_{0}>\omega_{\mathrm{LO}},$
the dominant part of the optical conductivity spectrum shifts to higher
frequencies. For instance, at $\Omega_{0}=1.4\omega_{\mathrm{LO}}$ the most
intensive peak is that with $\tilde{\Omega}_{3}$. The intensities of the
peaks, beginning with the second peak, increase in comparison with the
intensities of their ``weak-confinement'' analogs. The positions of the
zero-phonon line and the subsequent peaks are substantially shifted from the
``weak-confinement'' values towards higher frequencies. These effects are a
manifestation of the mixing of the zero-phonon state with different excited
states of the many-polaron system. A similar behavior of the optical
absorption spectra at and above the magneto-phonon resonance is explained by
the mixing of zero-phonon and one-phonon quantum states \cite{PRL66-794}.

The shell structure for a system of interacting polarons in a quantum dot is
clearly revealed when analyzing the addition energy and the first frequency
moment of the optical conductivity in parallel. In Figs.~\thinspace7 and 8, we
show both the function
\begin{equation}
\Theta\left(  N\right)  \equiv\left.  \left\langle \omega\right\rangle
\right|  _{N+1}-2\left.  \left\langle \omega\right\rangle \right|
_{N}+\left.  \left\langle \omega\right\rangle \right|  _{N-1},\label{Theta}%
\end{equation}
and the addition energy $\Delta\left(  N\right)  $ for interacting polarons in
different 3D quantum dots.

As seen from Figs.\thinspace7 and 8 for quantum dots of CdSe and with
$\alpha=3$\cite{alpha}, respectively, distinct peaks appear in $\Theta\left(
N\right)  $ and $\Delta\left(  N\right)  $ at the ``magic numbers''
corresponding to closed-shell configurations at $N=8,20$ for the states
obeying Hund's rule in panels $a,b$ and to half-filled-shell configurations at
$N=10,20$ for the spin-polarized states in panels $c,d$ of Fig.\thinspace8. In
the case when the shell filling scheme is the same for different $N$ (see
panels $a,b$ in Figs.\thinspace7,\thinspace8, where the filling obeys Hund's
rule), each of the peaks of $\Theta\left(  N\right)  $ corresponds to a peak
of the addition energy. In the case when the shell filling scheme changes with
varying $N$ (panels $c,d$ in Figs.\thinspace7,\thinspace8), the function
$\Theta\left(  N\right)  $ exhibits pronounced minima for $N$ corresponding to
the change of the filling scheme from the states, obeying Hund's rule, to the
spin-polarized states.

It follows that measurements of the addition energy and the first frequency
moment of the optical absorption as a function of the number of polarons in a
quantum dot can reflect the difference between open-shell and closed-shell
configurations. In particular, the closed-shell configurations may be revealed
through peaks in the function $\Theta\left(  N\right)  $. The filling patterns
for a many-polaron system in a quantum dot can be determined from the analysis
of the first moment of the optical absorption for different numbers of
polarons. The appearance of minima in the function $\Theta\left(  N\right)  $
will then indicate a transition from the states which are filled according to
Hund's rule to the spin-polarized states.\bigskip

\section{Conclusions}

\label{sec:conclusions}

We presented a formalism for calculating the ground-state energy and the
optical conductivity spectra of a system of $N$ interacting polarons in a
parabolic confinement potential for arbitrary electron-phonon coupling
strength. The path integral treatment of the quantum statistics of
indistinguishable particles \cite{PRE96,PRE97} allows us to find an upper
bound \cite{PRE96} to the ground-state energy of a finite number of polarons.
The parameters from the variational procedure are used as input for the
calculation of the optical conductivity spectrum of the system.

Two types of transitions were found for $N$ polarons confined in a parabolic
potential, with the corresponding ground states characterized by different
values of the total spin. In the weak-confinement regime, the polaron system
is in the spin-polarized state. When increasing the confinement frequency
$\Omega_{0}$, the system goes into a state obeying Hund's rule at a specific
value of $\Omega_{0}$. For a strongly coupled system of interacting polarons,
a third type of state appears, for which the total spin takes its minimal
value. The analysis is performed for both closed-shell and open-shell systems.

The calculations of the optical conductivity spectra for a finite number of
polarons in a quantum dot are based on the memory-function approach. The
dependence of the optical conductivity spectra on the confinement parameter
$\Omega_{0}$ reveals a resonant behavior for $\Omega_{0}\approx\omega
_{\mathrm{LO}}$. Transitions between states with different values of the total
spin manifest themselves through discontinuous changes of the optical
conductivity spectra and of the addition energy as a function of the number of electrons.

The first frequency moment of the optical conductivity as a function of the
number of electrons clearly shows the transition between the spin-polarized
ground state of interacting polarons in a quantum dot and the ground state
obeying Hund's rule, and it also can be used to discriminate between
open-shell and closed-shell configurations. Optical measurements are therefore
suggested as possible tools for examining the shell structure of a system of
interacting polarons.

\begin{acknowledgments}%

Discussions with V. N. Gladilin are gratefully
acknowledged. This work has been supported by the GOA BOF UA 2000, IUAP,
FWO-V projects G.0306.00, G.0274.01N, G.0435.03, the WOG WO.025.99N (Belgium)
and the European Commission GROWTH Programme, NANOMAT project, contract No.
G5RD-CT-2001-00545.
\end{acknowledgments}%

This work is performed in the framework of the FWO-V projects Nos. G.0435.03,
G.0306.00, the W.O.G. project WO.025.99N, and the GOA BOF UA 2000 project.

\appendix 

\section*{A. Partition function of the model system}%

\def\theequation{A.\arabic{equation}}
\setcounter{equation}{0}%

In this appendix we discuss the analytical calculation of the partition
function $Z_{0}\left(  \left\{  N_{\sigma}\right\}  ,\beta\right)  $ [Eq.
(\ref{Z0})] for the model system of interacting polarons. It can be expressed
in terms of the partition function $Z_{M}\left(  \left\{  N_{\sigma}\right\}
,N_{f},\beta\right)  $ of the model system of interacting electrons and
fictitious particles with the Lagrangian $L_{M}$ [Eq. (\ref{LM})] as follows:%
\begin{equation}
Z_{0}\left(  \left\{  N_{\sigma}\right\}  ,\beta\right)  =\frac{Z_{M}\left(
\left\{  N_{\sigma}\right\}  ,N_{f},\beta\right)  }{Z_{f}\left(  N_{f}%
,w_{f},\beta\right)  },\label{r1}%
\end{equation}
where $Z_{f}\left(  N_{f},w_{f},\beta\right)  $ is the partition function of
fictitious particles,%
\begin{equation}
Z_{f}\left(  N_{f},\beta\right)  =\frac{1}{\left(  2\sinh\frac{1}{2}\beta\hbar
w_{f}\right)  ^{DN_{f}}},\label{Zf}%
\end{equation}
with%
\begin{equation}
w_{f}=\sqrt{\Omega_{f}^{2}+kN/m_{f}}\label{wf}%
\end{equation}
and D=3(2) for 3D(2D) systems. The partition function $Z_{M}\left(  \left\{
N_{\sigma}\right\}  ,N_{f},\beta\right)  $ is the path integral for both the
electrons and the fictitious particles:%
\begin{align}
Z_{M}\left(  \left\{  N_{\sigma}\right\}  ,N_{f},\beta\right)   & =\sum
_{P}\frac{\left(  -1\right)  ^{\mathbf{\xi}_{P}}}{N_{1/2}!N_{-1/2}%
!}\nonumber\\
& \int d\mathbf{\bar{x}}\int_{\mathbf{\bar{x}}}^{P\mathbf{\bar{x}}%
}D\mathbf{\bar{x}}\left(  \tau\right)  \int d\mathbf{\bar{y}}\int
_{\mathbf{\bar{y}}}^{\mathbf{\bar{y}}}D\mathbf{\bar{y}}\left(  \tau\right)
e^{-S_{M}\left[  \mathbf{\bar{x}}\left(  \tau\right)  ,\mathbf{\bar{y}}\left(
\tau\right)  \right]  }\label{ZM}%
\end{align}
with the ``action'' functional
\begin{equation}
S_{M}\left[  \mathbf{\bar{x}}\left(  \tau\right)  ,\mathbf{\bar{y}}\left(
\tau\right)  \right]  =-\frac{1}{\hbar}\int_{0}^{\hbar\beta}L_{M}\left(
\mathbf{\dot{\bar{x}}},\mathbf{\dot{\bar{y}}};\mathbf{\bar{x}},\mathbf{\bar
{y}}\right)  d\tau,\label{SM}%
\end{equation}
where the Lagrangian is given by Eq. (\ref{LM}).

Let us consider an auxiliary ``ghost'' subsystem with the Lagrangian
\begin{equation}
L_{g}\left(  \mathbf{\dot{X}}_{g},\mathbf{\dot{Y}}_{g},\mathbf{X}%
_{g},\mathbf{Y}_{g}\right)  =-\frac{m_{b}N}{2}\left(  \mathbf{\dot{X}}_{g}%
^{2}+w^{2}\mathbf{X}_{g}^{2}\right)  -\frac{m_{f}N_{f}}{2}\left(
\mathbf{\dot{Y}}_{g}^{2}+w_{f}^{2}\mathbf{Y}_{g}^{2}\right) \label{ghost}%
\end{equation}
with two frequencies $w$ and $w_{f},$ where $w$ is given by%
\begin{equation}
w=\sqrt{\Omega^{2}-N\omega^{2}+kN_{f}/m_{b}}.\label{wab}%
\end{equation}
The partition function $Z_{g}$ of this subsystem
\begin{equation}
Z_{g}=\int d\mathbf{X}_{g}\int d\mathbf{Y}_{g}\int\limits_{\mathbf{X}_{g}%
}^{\mathbf{X}_{g}}D\mathbf{X}_{g}\left(  \tau\right)  \int\limits_{\mathbf{Y}%
_{g}}^{\mathbf{Y}_{g}}D\mathbf{Y}_{g}\left(  \tau\right)  \exp\left\{
-S_{g}\left[  \mathbf{X}_{g}\left(  \tau\right)  ,\mathbf{Y}_{g}\left(
\tau\right)  \right]  \right\}  ,\label{Zg1}%
\end{equation}
with the ``action'' functional%
\begin{equation}
S_{g}\left[  \mathbf{X}_{g}\left(  \tau\right)  ,\mathbf{Y}_{g}\left(
\tau\right)  \right]  =-\frac{1}{\hbar}\int\limits_{0}^{\hbar\beta}%
L_{g}\left(  \mathbf{\dot{X}}_{g},\mathbf{X}_{g},\mathbf{\dot{Y}}%
_{g},\mathbf{Y}_{g}\right)  \,d\tau\label{Sg}%
\end{equation}
is readily calculated:
\begin{equation}
Z_{g}=\frac{1}{\left[  2\sinh\left(  \frac{\beta\hbar w}{2}\right)  \right]
^{D}}\frac{1}{\left[  2\sinh\left(  \frac{\beta\hbar w_{f}}{2}\right)
\right]  ^{D}}.\label{Zg2}%
\end{equation}

The product $Z_{g}Z_{M}$ of the two partition functions $Z_{g}$ and
$Z_{M}\left(  \left\{  N_{\sigma}\right\}  ,N_{f},\beta\right)  $ is a path
integral in the state space of $N$ electrons, $N_{f}$ fictitious particles and
two ``ghost'' particles with the coordinate vectors $\mathbf{X}_{g}$ and
$\mathbf{Y}_{g}.$ The Lagrangian $\tilde{L}_{M}$ of this system is a sum of
$L_{M}$ and $L_{g},$%
\begin{equation}
\tilde{L}_{M}\left(  \mathbf{\dot{\bar{x}}},\mathbf{\dot{\bar{y}},\dot{X}}%
_{g},\mathbf{\dot{Y}}_{g};\mathbf{\bar{x}},\mathbf{\bar{y},X}_{g}%
,\mathbf{Y}_{g}\right)  \equiv L_{M}\left(  \mathbf{\dot{\bar{x}}%
},\mathbf{\dot{\bar{y}}};\mathbf{\bar{x}},\mathbf{\bar{y}}\right)
+L_{g}\left(  \mathbf{\dot{X}}_{g},\mathbf{\dot{Y}}_{g},\mathbf{X}%
_{g},\mathbf{Y}_{g}\right)  .\label{LMt}%
\end{equation}
The ``ghost'' subsystem is introduced because the center-of-mass coordinates
in $\tilde{L}_{M}$ can be explicitly separated more easily than in $L_{M}$.
This separation is realized by the linear transformation of coordinates,%
\begin{equation}
\left\{
\begin{array}
[c]{l}%
\mathbf{x}_{j,\sigma}=\mathbf{x}_{j,\sigma}^{\prime}+\mathbf{X}-\mathbf{X}%
_{g},\\
\mathbf{y}_{j\sigma}=\mathbf{y}_{j\sigma}^{\prime}+\mathbf{Y}-\mathbf{Y}_{g},
\end{array}
\right. \label{Trans}%
\end{equation}
where $\mathbf{X}$\ and $\mathbf{Y}$ are the center-of-mass coordinate vectors
of the electrons and of the fictitious particles, correspondingly:%
\begin{equation}
\mathbf{X=}\frac{1}{N}\sum_{\sigma}\sum_{j=1}^{N_{\sigma}}\mathbf{x}%
_{j,\sigma},\quad\mathbf{Y=}\frac{1}{N_{f}}\sum_{j=1}^{N_{f}}\mathbf{y}%
_{j}.\label{XY}%
\end{equation}
Before the transformation (\ref{Trans}), the independent variables are
$\left(  \mathbf{\bar{x}},\mathbf{\bar{y},X}_{g},\mathbf{Y}_{g}\right)  ,$
with the center-of-mass coordinates $\mathbf{X}$ and $\mathbf{Y}$ determined
by Eq. (\ref{XY}). After this transformation the independent variables can be
considered to be $\left(  \mathbf{\bar{x}}^{\prime},\mathbf{\bar{y}}^{\prime
},\mathbf{X,Y}\right)  ,$ where the coordinates $\left(  \mathbf{X}%
_{g},\mathbf{Y}_{g}\right)  $ obey the equations%
\begin{equation}
\mathbf{X}_{g}\mathbf{=}\frac{1}{N}\sum_{\sigma}\sum_{j=1}^{N_{\sigma}%
}\mathbf{x}_{j,\sigma}^{\prime},\quad\mathbf{Y}_{g}\mathbf{=}\frac{1}{N_{f}%
}\sum_{j=1}^{N_{f}}\mathbf{y}_{j}^{\prime}.\label{XgYg}%
\end{equation}
\qquad\qquad

A substitution of Eq. (\ref{XY}) into Eq. (\ref{LMt}) results in the following
3 terms:%
\begin{equation}
\tilde{L}_{M}\left(  \mathbf{\dot{\bar{x}}}^{\prime},\mathbf{\dot{\bar{y}}%
}^{\prime},\mathbf{\dot{X},\dot{Y};\bar{x}}^{\prime},\mathbf{\bar{y}}^{\prime
},\mathbf{X,Y}\right)  =L_{w}\left(  \mathbf{\dot{\bar{x}}}^{\prime
},\mathbf{\bar{x}}^{\prime}\right)  +L_{w_{f}}\left(  \mathbf{\dot{\bar{y}}%
}^{\prime},\mathbf{\bar{y}}^{\prime}\right)  +L_{C}\left(  \mathbf{\dot{X}%
,X};\mathbf{\dot{Y},Y}\right)  ,\label{LM1}%
\end{equation}
where $L_{w}\left(  \mathbf{\dot{\bar{x}}}^{\prime},\mathbf{\bar{x}}^{\prime
}\right)  $ and $L_{w_{f}}\left(  \mathbf{\dot{\bar{y}}}^{\prime}%
,\mathbf{\bar{y}}^{\prime}\right)  $ are Lagrangians of non-interacting
identical oscillators with the frequencies $w$ and $w_{f},$ respectively,%
\begin{align}
L_{w}\left(  \mathbf{\dot{\bar{x}}}^{\prime},\mathbf{\bar{x}}^{\prime
}\right)   & =-\frac{m_{b}}{2}\sum_{\sigma=\pm1/2}\sum_{j=1}^{N_{\sigma}%
}\left[  \left(  \mathbf{\dot{x}}_{j,\sigma}^{\prime}\right)  ^{2}%
+w^{2}\left(  \mathbf{x}_{j,\sigma}^{\prime}\right)  ^{2}\right]
,\label{Lwa}\\
L_{w_{f}}\left(  \mathbf{\dot{\bar{y}}}^{\prime},\mathbf{\bar{y}}^{\prime
}\right)   & =-\frac{m_{f}}{2}\sum_{j=1}^{N_{f}}\left[  \left(  \mathbf{\dot
{y}}_{j,\sigma}^{\prime}\right)  ^{2}+w_{f}^{2}\left(  \mathbf{y}_{j,\sigma
}^{\prime}\right)  ^{2}\right]  .\label{Lwb}%
\end{align}
The Lagrangian $L_{C}\left(  \mathbf{\dot{X},X};\mathbf{\dot{Y},Y}\right)  $
describes the combined motion of the centers-of-mass of the electrons and of
the fictitious particles,%
\begin{equation}
L_{C}\left(  \mathbf{\dot{X},X};\mathbf{\dot{Y},Y}\right)  =-\frac{m_{b}N}%
{2}\left(  \mathbf{\dot{X}}^{2}+\tilde{\Omega}^{2}\mathbf{X}^{2}\right)
-\frac{m_{f}N_{f}}{2}\left(  \mathbf{\dot{Y}}^{2}+w_{f}^{2}\mathbf{Y}%
^{2}\right)  +kNN_{f}\mathbf{X\cdot Y,}\label{LC}%
\end{equation}
with%
\begin{equation}
\tilde{\Omega}=\sqrt{\Omega^{2}+kN_{f}/m_{b}}.\label{Ot}%
\end{equation}
It is reduced to a diagonal quadratic form in the coordinates and the
velocities by a standard transformation for two interacting oscillators:
\begin{align}
\mathbf{X}  & =\frac{1}{\sqrt{m_{b}N}}\left(  a_{1}\mathbf{r}+a_{2}%
\mathbf{R}\right)  ,\nonumber\\
\mathbf{Y}  & =\frac{1}{\sqrt{m_{f}N_{f}}}\left(  -a_{2}\mathbf{r}%
+a_{1}\mathbf{R}\right)
\end{align}
with the coefficients
\begin{align}
a_{1}  & =\left[  \frac{1+\chi}{2}\right]  ^{1/2},\quad a_{2}=\left[
\frac{1-\chi}{2}\right]  ^{1/2},\label{a12}\\
\chi & \equiv\frac{\tilde{\Omega}^{2}-\tilde{\Omega}_{f}^{2}}{\left[  \left(
\tilde{\Omega}^{2}-\tilde{\Omega}_{f}^{2}\right)  ^{2}+4\gamma^{2}\right]
^{1/2}},\quad\gamma\equiv k\sqrt{\frac{NN_{f}}{m_{b}m_{f}}}.\label{C}%
\end{align}
The eigenfrequencies of the center-of-mass subsystem are then given by the
expression
\begin{equation}
\left\{
\begin{array}
[c]{c}%
\Omega_{1}=\sqrt{\frac{1}{2}\left[  \tilde{\Omega}^{2}+\tilde{\Omega}_{f}%
^{2}+\sqrt{\left(  \tilde{\Omega}^{2}-\tilde{\Omega}_{f}^{2}\right)
^{2}+4\gamma^{2}}\right]  },\\
\Omega_{2}=\sqrt{\frac{1}{2}\left[  \tilde{\Omega}^{2}+\tilde{\Omega}_{f}%
^{2}-\sqrt{\left(  \tilde{\Omega}^{2}-\tilde{\Omega}_{f}^{2}\right)
^{2}+4\gamma^{2}}\right]  }.
\end{array}
\right. \label{O12}%
\end{equation}
As a result, four independent frequencies $\Omega_{1},$ $\Omega_{2},$ $w$ and
$w_{f}$ appear in the problem. Three of them ($\Omega_{1},$ $\Omega_{2},$ $w$)
are the eigenfrequencies of the model system. $\Omega_{1}$\ is the frequency
of the relative motion of the center of mass of the electrons with respect to
the center of mass of the fictitious particles; $\Omega_{2}$\ is the frequency
related to the center of mass of the model system as a whole; $w$\ is the
frequency of the relative motion of the electrons with respect to their center
of mass. The parameter $w_{f}$ is an analog of the second variational
parameter $w$ of the one-polaron Feynman model. Further, the Lagrangian
(\ref{LC}) takes the form
\begin{equation}
L_{C}=-\frac{1}{2}\left(  \mathbf{\dot{r}}^{2}+\Omega_{1}^{2}\mathbf{r}%
^{2}\right)  -\frac{1}{2}\left(  \mathbf{\dot{R}}^{2}+\Omega_{2}^{2}%
\mathbf{R}^{2}\right)  ,\label{QF}%
\end{equation}
leading to the partition function corresponding to the combined motion of the
centers-of-mass of the electrons and of the fictitious particles
\begin{equation}
Z_{C}=\frac{1}{\left[  2\sinh\left(  \frac{\beta\hbar\Omega_{1}}{2}\right)
\right]  ^{D}}\frac{1}{\left[  2\sinh\left(  \frac{\beta\hbar\Omega_{2}}%
{2}\right)  \right]  ^{D}}.\label{ZC}%
\end{equation}

Taking into account Eqs. (\ref{Zg2}) and (\ref{ZC}), we obtain finally the
partition function of the model system for interacting polarons
\begin{equation}
Z_{0}\left(  \left\{  N_{\sigma}\right\}  ,\beta\right)  =\left[
\frac{\sinh\left(  \frac{\beta\hbar w}{2}\right)  \sinh\left(  \frac{\beta
\hbar w_{f}}{2}\right)  }{\sinh\left(  \frac{\beta\hbar\Omega_{1}}{2}\right)
\sinh\left(  \frac{\beta\hbar\Omega_{2}}{2}\right)  }\right]  ^{D}%
\mathbb{\tilde{Z}}_{F}\left(  \left\{  N_{\sigma}\right\}  ,w,\beta\right)
.\label{Z02}%
\end{equation}
Here%
\begin{equation}
\mathbb{\tilde{Z}}_{F}\left(  \left\{  N_{\sigma}\right\}  ,w,\beta\right)
=\mathbb{Z}_{F}\left(  N_{1/2},w,\beta\right)  \mathbb{Z}_{F}\left(
N_{-1/2},w,\beta\right) \label{sup}%
\end{equation}
is the partition function of $N=N_{1/2}+N_{-1/2}$ non-interacting fermions in
a parabolic confinement potential with the frequency $w.$ The analytical
expressions for the partition function of $N_{\sigma}$ spin-polarized fermions
$\mathbb{Z}_{F}\left(  N_{\sigma},w,\beta\right)  $ were derived in Ref.
\cite{PRE97}.

\section*{B. Two-point correlation function. Memory function.}%

\def\theequation{B.\arabic{equation}}
\setcounter{equation}{0}%

The two-point correlation function (\ref{TPCF}) is represented as the
following path integral:%
\begin{align}
g\left(  \mathbf{q},\tau|\left\{  N_{\sigma}\right\}  ,\beta\right)   &
=\frac{1}{Z_{0}\left(  \left\{  N_{\sigma}\right\}  ,\beta\right)  }\sum
_{P}\frac{\left(  -1\right)  ^{\mathbf{\xi}_{P}}}{N_{1/2}!N_{-1/2}%
!}\nonumber\\
& \int d\mathbf{\bar{x}}\int_{\mathbf{\bar{x}}}^{P\mathbf{\bar{x}}%
}D\mathbf{\bar{x}}\left(  \tau\right)  e^{-S_{0}\left[  \mathbf{\bar{x}%
}\left(  \tau\right)  \right]  }\rho_{\mathbf{q}}\left(  \tau\right)
\rho_{-\mathbf{q}}\left(  0\right)  .\label{g1}%
\end{align}
We observe that $g\left(  \mathbf{q},\tau|\left\{  N_{\sigma}\right\}
,\beta\right)  $ can be rewritten as an average within the model ``action''
$S_{M}\left[  \mathbf{\bar{x}}\left(  \tau\right)  ,\mathbf{\bar{y}}\left(
\tau\right)  \right]  $ of interacting electrons and fictitious particles:%
\begin{align}
g\left(  \mathbf{q},\tau|\left\{  N_{\sigma}\right\}  ,\beta\right)   &
=\frac{1}{Z_{M}\left(  \left\{  N_{\sigma}\right\}  ,N_{f},\beta\right)  }%
\sum_{P}\frac{\left(  -1\right)  ^{\mathbf{\xi}_{P}}}{N_{1/2}!N_{-1/2}%
!}\nonumber\\
& \times\int d\mathbf{\bar{x}}\int_{\mathbf{\bar{x}}}^{P\mathbf{\bar{x}}%
}D\mathbf{\bar{x}}\left(  \tau\right)  \int d\mathbf{\bar{y}}\int
_{\mathbf{\bar{y}}}^{\mathbf{\bar{y}}}D\mathbf{\bar{y}}\left(  \tau\right)
e^{-S_{M}\left[  \mathbf{\bar{x}}\left(  \tau\right)  ,\mathbf{\bar{y}}\left(
\tau\right)  \right]  }\nonumber\\
& \times\rho_{\mathbf{q}}\left(  \tau\right)  \rho_{-\mathbf{q}}\left(
0\right)  .\label{g2}%
\end{align}
Indeed, one readily derives that the elimination of the fictitious particles
in (\ref{g2}) leads to (\ref{g1}). The representation (\ref{g2}) allows one to
calculate the correlation function $g\left(  \mathbf{q},\tau|\left\{
N_{\sigma}\right\}  ,\beta\right)  $ in a much simpler way than through Eq.
(\ref{g1}), using the separation of the coordinates of the centers of mass of
the electrons and of the fictitious particles. This separation is performed
for the two-point correlation function (\ref{g2}) by the same method as it has
been done for the partition function (\ref{ZM}). As a result, one obtains%
\begin{equation}
g\left(  \mathbf{q},\tau|\left\{  N_{\sigma}\right\}  ,\beta\right)
=\tilde{g}\left(  \mathbf{q},\tau|\left\{  N_{\sigma}\right\}  ,\beta\right)
\frac{\left\langle \exp\left[  i\mathbf{q\cdot}\left(  \mathbf{X}\left(
\tau\right)  -\mathbf{X}\left(  \sigma\right)  \right)  \right]  \right\rangle
_{S_{C}}}{\left\langle \exp\left[  i\mathbf{q\cdot}\left(  \mathbf{X}%
_{g}\left(  \tau\right)  -\mathbf{X}_{g}\left(  \sigma\right)  \right)
\right]  \right\rangle _{S_{g}}},\label{Fact2}%
\end{equation}
where $\tilde{g}\left(  \mathbf{q},\tau|\left\{  N_{\sigma}\right\}
,\beta\right)  $ is the time-dependent correlation function of $N$
non-interacting electrons in a parabolic confinement potential with the
frequency $w$,
\begin{equation}
\tilde{g}\left(  \mathbf{q},\tau|\left\{  N_{\sigma}\right\}  ,\beta\right)
=\left\langle \rho_{\mathbf{q}}\left(  \tau\right)  \rho_{-\mathbf{q}}\left(
0\right)  \right\rangle _{S_{w}}.\label{Gqtild}%
\end{equation}
The action functional $S_{w}\left[  \mathbf{\bar{x}}_{\tau}\right]  $ is
related to the Lagrangian $L_{w}\left(  \mathbf{\dot{\bar{x}}},\mathbf{\bar
{x}}\right)  $ [Eq. (\ref{Lwa})]%
\begin{equation}
S_{w}\left[  \mathbf{\bar{x}}_{\tau}\right]  =\frac{1}{\hbar}\int
\limits_{0}^{\hbar\beta}L_{w}\left(  \mathbf{\dot{\bar{x}}},\mathbf{\bar{x}%
}\right)  \,d\tau.\label{Sw}%
\end{equation}
The averages in (\ref{Fact2}) are calculated using Feynman's method of
generating functions \cite{Feynman},
\begin{align*}
\left\langle \exp\left[  i\mathbf{q\cdot}\left(  \mathbf{X}\left(
\tau\right)  -\mathbf{X}\left(  \sigma\right)  \right)  \right]  \right\rangle
_{S_{C}}  & =\exp\left\{  -\frac{\hbar q^{2}}{Nm_{b}}\left[  \sum_{i=1}%
^{2}a_{i}^{2}\frac{\sinh\left(  \frac{\Omega_{i}\left|  \tau-\sigma\right|
}{2}\right)  \sinh\left(  \frac{\Omega_{i}\left(  \hbar\beta-\left|
\tau-\sigma\right|  \right)  }{2}\right)  }{\Omega_{i}\sinh\left(
\frac{\beta\hbar\Omega_{i}}{2}\right)  }\right]  \right\}  ,\\
\left\langle \exp\left[  i\mathbf{q\cdot}\left(  \mathbf{X}_{g}\left(
\tau\right)  -\mathbf{X}_{g}\left(  \sigma\right)  \right)  \right]
\right\rangle _{S_{g}}  & =\exp\left[  -\frac{\hbar q^{2}}{Nm_{b}}%
\frac{\sinh\left(  \frac{w\left|  \tau-\sigma\right|  }{2}\right)
\sinh\left(  \frac{w\left(  \hbar\beta-\left|  \tau-\sigma\right|  \right)
}{2}\right)  }{w\sinh\left(  \frac{\beta\hbar w}{2}\right)  }\right]  .
\end{align*}
The two-point correlation function $\tilde{g}\left(  \mathbf{q},\tau|\left\{
N_{\sigma}\right\}  ,\beta\right)  $ is derived using the generating-function
technique for identical particles \cite{PRE97}. After the path integration,
the following expression is obtained:%
\begin{align}
& \tilde{g}\left(  \mathbf{q},-i\tau|\left\{  N_{\sigma}\right\}
,\beta\right) \nonumber\\
& =\sum_{\mathbf{n},\sigma,\mathbf{n}^{\prime},\sigma^{\prime}}\left(
e^{i\mathbf{q\cdot x}}\right)  _{\mathbf{nn}}\left(  e^{-i\mathbf{q\cdot x}%
}\right)  _{\mathbf{n^{\prime}n^{\prime}}}f_{2}\left(  n,\sigma;n^{\prime
},\sigma^{\prime}|\left\{  N_{\sigma}\right\}  ,\beta\right) \nonumber\\
& +\sum_{\mathbf{n},\mathbf{n}^{\prime},\sigma}\left|  \left(
e^{i\mathbf{q\cdot x}}\right)  _{\mathbf{nn^{\prime}}}\right|  ^{2}\exp\left[
\frac{\tau}{\hbar}\left(  \varepsilon_{n}-\varepsilon_{n^{\prime}}\right)
\right] \nonumber\\
& \times\left[  f_{1}\left(  n,\sigma|\left\{  N_{\sigma}\right\}
,\beta\right)  -f_{2}\left(  n,\sigma;n^{\prime},\sigma|\left\{  N_{\sigma
}\right\}  ,\beta\right)  \right]  ,\label{g4}%
\end{align}
where $\left(  e^{i\mathbf{q\cdot x}}\right)  _{\mathbf{nn^{\prime}}}$ is the
one-electron matrix element,%
\begin{equation}
\left(  e^{i\mathbf{q\cdot x}}\right)  _{\mathbf{nn^{\prime}}}=\int
e^{i\mathbf{q\cdot x}}\psi_{\mathbf{n}}^{\ast}\left(  \mathbf{x}\right)
\psi_{\mathbf{n}^{\prime}}\left(  \mathbf{x}\right)  d\mathbf{x.}\label{me}%
\end{equation}
For a 3D quantum dot, $\psi_{\mathbf{n}}\left(  \mathbf{x}\right)  $ is the
eigenfunction of a 3D oscillator with the frequency $w$ [see, e.g., Ref.
\cite{Cohen}]. The index $\mathbf{n}$ denotes the set $\mathbf{n}=\left(
n,l,m\right)  ,$ where $n$ is the number of the energy level $\varepsilon
_{n}=\hbar w\left(  n+3/2\right)  $, $l$ is the quantum number of the orbital
angular momentum and $m$ is the quantum number of the $z$ projection of the
orbital angular momentum. Similarly, for a 2D quantum dot, $\psi_{\mathbf{n}%
}\left(  \mathbf{x}\right)  $ is the eigenfunction of a 2D oscillator with the
frequency $w.$

The one-electron distribution function $f_{1}\left(  n,\sigma|N_{\sigma}%
,\beta\right)  $ is the average number of electrons with the spin projection
$\sigma$ at the $n$-th energy level, while the two-electron distribution
function $f_{2}\left(  n,\sigma;n^{\prime},\sigma^{\prime}|\left\{  N_{\sigma
}\right\}  ,\beta\right)  $ is the average product of the numbers of electrons
with the spin projections $\sigma$ and $\sigma^{\prime}$ at the levels $n$ and
$n^{\prime}.$ These functions are expressed through the following integrals
(cf. Ref. \cite{SSC99}):%
\begin{align}
f_{1}\left(  n,\sigma|N_{\sigma},\beta\right)   & =\frac{1}{2\pi\mathbb{Z}%
_{F}\left(  N_{\sigma},w,\beta\right)  }\int\limits_{-\pi}^{\pi}f\left(
\varepsilon_{n},\theta\right)  \Phi\left(  \theta,\beta,N_{\sigma}\right)
d\theta,\label{Num2a}\\
f_{2}\left(  n,\sigma;n^{\prime},\sigma^{\prime}|\left\{  N_{\sigma}\right\}
,\beta\right)   & =\left\{
\begin{array}
[c]{c}%
\frac{1}{2\pi\mathbb{Z}_{F}\left(  N_{\sigma},w,\beta\right)  }\int
\limits_{-\pi}^{\pi}f\left(  \varepsilon_{n},\theta\right)  f\left(
\varepsilon_{n\mathbf{^{\prime}}},\theta\right)  \Phi\left(  \theta
,\beta,N_{\sigma}\right)  d\theta,\;\text{if }\sigma^{\prime}=\sigma;\\
f_{1}\left(  n,\sigma|N_{\sigma},\beta\right)  f_{1}\left(  n,\sigma^{\prime
}|N_{\sigma^{\prime}},\beta\right)  ,\;\text{if }\sigma^{\prime}\neq\sigma
\end{array}
\right. \label{Num2b}%
\end{align}
with the notations%
\begin{equation}
\Phi\left(  \theta,\beta,N_{\sigma}\right)  =\exp\left[  \sum_{n=0}^{\infty
}\ln\left(  1+e^{i\theta+\mathbf{\xi}-\beta\varepsilon_{n}}\right)
-N_{\sigma}\left(  \mathbf{\xi}+i\theta\right)  \right]  ,\label{FiN}%
\end{equation}%
\begin{equation}
f\left(  \varepsilon,\theta\right)  \equiv\frac{1}{\exp\left(  \beta
\varepsilon-\mathbf{\xi}-i\theta\right)  +1}.\label{ferm}%
\end{equation}
The function $f\left(  \varepsilon,\theta\right)  $ formally coincides with
the Fermi-Dirac distribution function of the energy $\varepsilon$ with the
``chemical potential'' $\left(  \mathbf{\xi}+i\theta\right)  /\beta.$

From here on we consider the zero-temperature limit, for which the integrals
(\ref{Num2a}) and (\ref{Num2b}) can be calculated analytically. The result for
the one-electron distribution function is%
\begin{equation}
\left.  f_{1}\left(  n,\sigma|\beta,N_{\sigma}\right)  \right|  _{\beta
\rightarrow\infty}=\left\{
\begin{array}
[c]{cc}%
1, & n<L_{\sigma};\\
0, & n>L_{\sigma};\\
\frac{N_{\sigma}-N_{L_{\sigma}}}{g_{L_{\sigma}}}, & n=L_{\sigma}.
\end{array}
\right. \label{p1}%
\end{equation}
According to (\ref{p1}), $L_{\sigma}$ is the number of the lowest open shell,
and
\[
g_{n}=\left\{
\begin{array}
[c]{ccc}%
\frac{1}{2}\left(  n+1\right)  \left(  n+2\right)  &  & \left(  3D\right)  ,\\
n+1 &  & \left(  2D\right)  .
\end{array}
\right.
\]
is the degeneracy of the $n$-th shell. $N_{L_{\sigma}}$ is the number of
electrons in all the closed shells with the spin projection $\sigma,$%
\begin{equation}
N_{L_{\sigma}}\equiv\sum_{n=0}^{L_{\sigma}-1}g_{n}=\left\{
\begin{array}
[c]{ccc}%
\frac{1}{6}L_{\sigma}\left(  L_{\sigma}+1\right)  \left(  L_{\sigma}+2\right)
&  & \left(  3D\right)  ,\\
\frac{1}{2}L_{\sigma}\left(  L_{\sigma}+1\right)  &  & \left(  2D\right)  .
\end{array}
\right. \label{NLs}%
\end{equation}
The two-electron distribution function $f_{2}\left(  n,\sigma;n^{\prime
},\sigma^{\prime}|\left\{  N_{\sigma}\right\}  ,\beta\right)  $ at $T=0$ takes
the form%
\begin{align}
& \left.  f_{2}\left(  n,\sigma;n^{\prime},\sigma^{\prime}|\beta,\left\{
N_{\sigma}\right\}  \right)  \right|  _{\beta\rightarrow\infty}\nonumber\\
& =\left\{
\begin{array}
[c]{cc}%
\left.  f_{1}\left(  n,\sigma|\beta,N_{\sigma}\right)  \right|  _{\beta
\rightarrow\infty}\left.  f_{1}\left(  n^{\prime},\sigma^{\prime}%
|\beta,N_{\sigma^{\prime}}\right)  \right|  _{\beta\rightarrow\infty}, & n\neq
n^{\prime}\;\mathrm{or}\;\sigma\neq\sigma^{\prime}\\
1, & \sigma=\sigma^{\prime}\;\mathrm{and}\;n=n^{\prime}<L_{\sigma};\\
0, & \sigma=\sigma^{\prime}\;\mathrm{and}\;n=n^{\prime}>L_{\sigma};\\
\frac{N-N_{L_{\sigma}}}{g_{L_{\sigma}}}\frac{N-N_{L_{\sigma}}-1}{g_{L_{\sigma
}}-1}, & \sigma=\sigma^{\prime}\;\mathrm{and}\;n=n^{\prime}=L_{\sigma}.
\end{array}
\right. \label{p2}%
\end{align}
Finally, using the two-point correlation function Eq.~ (\ref{me}), the
one-electron (\ref{p1}) and the two-electron (\ref{p2}) distribution
functions, the memory function of Eq.~(\ref{Hi}) can be represented in the
unified form for 3D and 2D interacting polarons:
\begin{align}
\chi\left(  \omega\right)   & =\lim_{\varepsilon\rightarrow+0}\frac{2\alpha
}{3\pi N}\left(  \frac{3\pi}{4}\right)  ^{3-D}\left(  \frac{\omega
_{\mathrm{LO}}}{A}\right)  ^{3/2}\nonumber\\
& \times\sum_{p_{1}=0}^{\infty}\sum_{p_{2}=0}^{\infty}\sum_{p_{3}=0}^{\infty
}\frac{\left(  -1\right)  ^{p_{3}}}{p_{1}!p_{2}!p_{3}!}\left(  \frac{a_{1}%
^{2}}{N\Omega_{1}A}\right)  ^{p_{1}}\left(  \frac{a_{2}^{2}}{N\Omega_{2}%
A}\right)  ^{p_{2}}\left(  \frac{1}{NwA}\right)  ^{p_{3}}\nonumber\\
& \times\left\{  \left[  \sum\limits_{m=0}^{\infty}\sum\limits_{n=0}^{\infty
}\sum\limits_{\sigma}\left.  \left[  f_{1}\left(  n,\sigma|\left\{  N_{\sigma
}\right\}  ,\beta\right)  -f_{2}\left(  n,\sigma;m,\sigma|\left\{  N_{\sigma
}\right\}  ,\beta\right)  \right]  \right|  _{\beta\rightarrow\infty}\right.
\right. \nonumber\\
& \times\left(
\begin{array}
[c]{c}%
\frac{1}{\omega-\omega_{\mathrm{LO}}-\left[  p_{1}\Omega_{1}+p_{2}\Omega
_{2}+\left(  p_{3}-m+n\right)  w\right]  +\mathrm{i}\varepsilon}%
-\frac{1}{\omega+\omega_{\mathrm{LO}}+p_{1}\Omega_{1}+p_{2}\Omega_{2}+\left(
p_{3}-m+n\right)  w+\mathrm{i}\varepsilon}\\
+\mathcal{P}\left(  \frac{2}{\omega_{\mathrm{LO}}+p_{1}\Omega_{1}+p_{2}%
\Omega_{2}+\left(  p_{3}-m+n\right)  w}\right)
\end{array}
\right) \nonumber\\
& \times\sum\limits_{l=0}^{m}\sum\limits_{k=n-m+l}^{n}\frac{\left(  -1\right)
^{n-m+l+k}\Gamma\left(  p_{1}+p_{2}+p_{3}+k+l+\frac{3}{2}\right)  }%
{k!l!}\left(  \frac{1}{wA}\right)  ^{l+k}\nonumber\\
& \left.  \times\binom{n+D-1}{n-k}\binom{2k}{k-l-n+m}\right] \nonumber\\
& +\left[  \left(
\begin{array}
[c]{c}%
\frac{1}{\omega-\omega_{\mathrm{LO}}-\left(  p_{1}\Omega_{1}+p_{2}\Omega
_{2}+p_{3}w\right)  +\mathrm{i}\varepsilon}-\frac{1}{\omega+\omega
_{\mathrm{LO}}+p_{1}\Omega_{1}+p_{2}\Omega_{2}+p_{3}w+\mathrm{i}\varepsilon}\\
+\mathcal{P}\left(  \frac{2}{\omega_{\mathrm{LO}}+p_{1}\Omega_{1}+p_{2}%
\Omega_{2}+p_{3}w}\right)
\end{array}
\right)  \right. \nonumber\\
& \times\sum\limits_{m=0}^{\infty}\sum\limits_{n=0}^{\infty}\sum
\limits_{\sigma,\sigma^{\prime}}\left.  f_{2}\left(  n,\sigma;m,\sigma
^{\prime}|\left\{  N_{\sigma}\right\}  ,\beta\right)  \right|  _{\beta
\rightarrow\infty}\nonumber\\
& \times\sum\limits_{k=0}^{n}\sum\limits_{l=0}^{m}\frac{\left(  -1\right)
^{k+l}\Gamma\left(  p_{1}+p_{2}+p_{3}+k+l+\frac{3}{2}\right)  }{k!l!}\left(
\frac{1}{wA}\right)  ^{k+l}\nonumber\\
& \left.  \left.  \times\binom{n+D-1}{n-k}\binom{m+D-1}{m-l}\right]  \right\}
,\label{mf}%
\end{align}
where $D=2,3$ is the dimensionality of the space, $\mathcal{P}$ denotes the
principal value, $A$ is defined as $A\equiv\left[  \sum_{i=1}^{2}a_{i}%
^{2}/\Omega_{i}+\left(  N-1\right)  /w\right]  /N$, $\Omega_{1},\Omega_{2},$
and $w$ are the eigenfrequencies of the model system, $a_{1}$ and $a_{2}$ are
the coefficients of the canonical transformation which diagonalizes the model
Lagrangian (\ref{LM}) derived in Appendix A.

The typical spectra of the real and imaginary parts of the memory function
$\chi\left(  \omega\right)  ,$ are plotted in Figs. 9(\emph{a}) and
9(\emph{b}), respectively. According to Eq. (\ref{mf}), the poles of
$\operatorname{Re}\chi\left(  \omega\right)  $ and the $\delta$-like peaks of
$\left[  -\operatorname{Im}\chi\left(  \omega\right)  \right]  $ are
positioned at the combinatorial frequencies $\omega_{klm}$, which are linear
combinations of the LO-phonon frequency and three eigenfrequencies%
\begin{equation}
\omega_{klm}\equiv\omega_{\mathrm{LO}}+k\Omega_{1}+l\Omega_{2}+mw,\label{oklm}%
\end{equation}
with integer coefficients $k,$ $l,$ $m=0,1,\ldots.$ Each combinatorial
frequency $\omega_{klm}$ corresponds to a phonon-assisted transition to an
excited state of the model system.

The roots of the equation (\ref{peaks}), which provide the peaks in the
optical conductivity, $\left(  \tilde{\Omega}_{1},\tilde{\Omega}_{2}%
,\ldots\right)  $, are indicated in Fig. 9\emph{a} by the vertical arrows. For
the chosen parameters, the peak at $\tilde{\Omega}_{1}$ is the zero-phonon
line. Fig. 9\emph{a} also reveals peaks of $\mathop{\rm Re} \sigma\left(
\omega\right)  $ with frequencies in between two neighboring discrete values
of $\omega_{nkl}$, e.g., at $\tilde{\Omega}_{2}$. Following the physical
interpretation of the memory function in Refs.~\cite{DSG,Green,Bipabs}, these
peaks can be related to transitions into excited states of the many-polaron system.

\newpage

Figure captions

\bigskip

Fig. 1. Total spin of a system of interacting polarons in a parabolic quantum
dot as a function of the number of electrons for $\hbar\Omega_{0}=0.5H^{\ast}%
$(\emph{a}) and for $\hbar\Omega_{0}=0.1H^{\ast}$(\emph{b}).

\medskip

Fig. 2. Addition energy of a system of interacting polarons in a parabolic
quantum dot as a function of the number of electrons for $\hbar\Omega
_{0}=0.5H^{\ast}$(\emph{a}) and for $\hbar\Omega_{0}=0.1H^{\ast}$(\emph{b}).

\medskip

Fig. 3. The total spin (\emph{a}) and the addition energy (\emph{b}) of a
system of interacting polarons in a 2D parabolic GaAs quantum dot as a
function of the number of electrons for $\hbar\Omega_{0}=0.5$$H^{\ast}$.
Inset: the experimentally observed addition energy vs number of electrons in a
cylindrical GaAs quantum dot for two values of the diameter $D=0.5\mu$m and
$D=0.44\mu$m \cite{Tarucha1996}.

\medskip

Fig. 4. Optical conductivity spectra of $N=20$ interacting polarons in CdSe
quantum dots with $\alpha=0.46$, $\eta=0.656$ for different confinement
energies close to the transition from a spin-polarized ground state to a
ground state obeying Hund$^{\text{'}}$s rule. \emph{Inset}: the first
frequency moment $\left\langle \omega\right\rangle $ of the optical
conductivity as a function of the confinement energy.

\medskip

Fig. 5. The first frequency moment $\left\langle \omega\right\rangle $of the
optical conductivity as a function of the number of electrons for systems of
interacting polarons in CdSe quantum dots with $\alpha=0.46$, $\eta=0.656$ and
$0.143\omega_{\mathrm{LO}}$ ($\hbar\Omega_{0}\approx0.04H^{\ast}$). Open
squares denote the spin-polarized ground state; full dots denote the ground
state, obeying Hund's rule; open triangles denote the ground state of the
third type, with more than one partly filled shells, which is not totally
spin-polarized. \emph{Inset}: the total spin of the system of interacting
polarons as a function of $N.$

\medskip

Fig. 6. Optical conductivity spectra of $N=10$ interacting polarons in a
quantum dot with $\alpha=2$, $\eta=0.6$ for several values of the confinement
frequency from $\Omega_{0}=0.6\omega_{\mathrm{LO}}$ to $\Omega_{0}%
=1.4\omega_{\mathrm{LO}}.$ The spectrum for $\Omega_{0}=\omega_{\mathrm{LO}} $
corresponds to the confinement-phonon resonance.

\medskip

Fig. 7. The function $\Theta\left(  N\right)  $ and the addition energy
$\Delta\left(  N\right)  $ for systems of interacting polarons in CdSe quantum
dots with $\alpha=0.46$, $\eta=0.656$ for $\hbar\Omega_{0}=0.1H^{\ast}$
(panels $a,b$) and for $\Omega_{0}=0.04H^{\ast}$ (panels $c,d$). Open squares
denote the spin-polarized ground state; full dots denote the ground state,
obeying Hund's rule; open triangles denote the ground state of the third type,
with more than one partly filled shells, which is not totally spin-polarized.

\bigskip

Fig. 8. The function $\Theta\left(  N\right)  $ and the addition energy
$\Delta\left(  N\right)  $ for systems of interacting polarons in quantum dots
with $\alpha=3,$ $\eta=0.25$ and $\Omega_{0}=\omega_{\mathrm{LO}}$ (panels
$a,b$) and with $\alpha=3,$ $\eta=0.3$ and $\Omega_{0}=0.5\omega_{\mathrm{LO}%
}$ (panels $c,d$).

\bigskip

Fig. 9. Real (\emph{a}) and imaginary (\emph{b}) parts of the memory function
$\chi\left(  \omega\right)  /\omega$ for a system of interacting polarons in a
quantum dot for $N=10,$ $\alpha=2,$ $\eta=0.6,$ and $\Omega_{0}=0.6\omega
_{\mathrm{LO}}$. The dashed line in panel \emph{a }represents $\left(
\omega^{2}-\Omega_{0}^{2}\right)  /\omega.$ The vertical arrows in panel
\emph{a} indicate the roots of Eq.~(\ref{peaks}). The height of peaks in panel
\emph{b} represents the relative intensity of the $\delta$-like peaks of
$\left[  -\mathrm{Im}\chi\left(  \omega\right)  /\omega\right]  $.

\newpage%

\begin{figure}
[b]
\begin{center}
\includegraphics[
height=4.7288in,
width=3.1151in
]%
{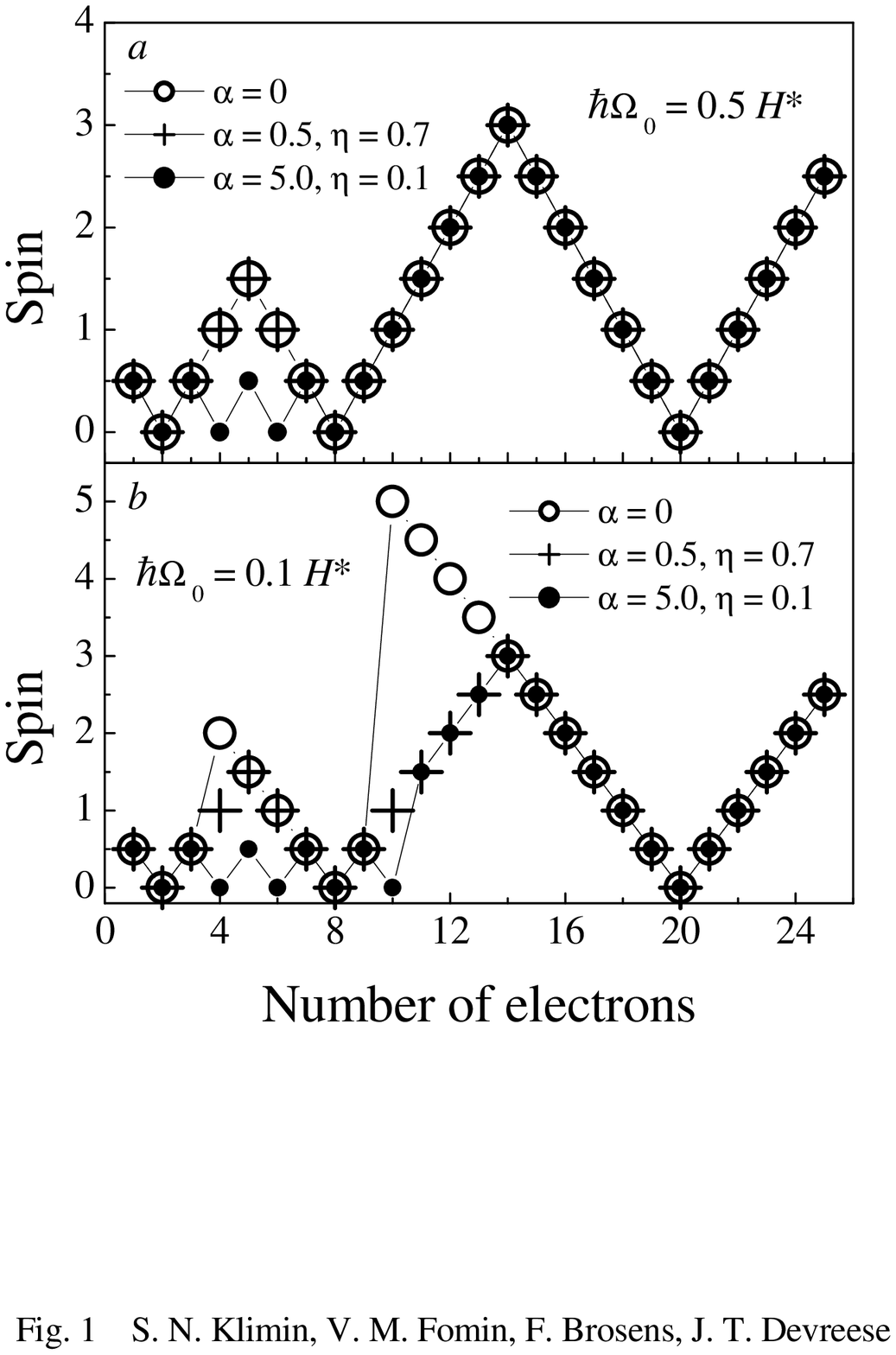}%
\end{center}
\end{figure}

\newpage%

\begin{figure}
[b]
\begin{center}
\includegraphics[
height=4.6458in,
width=3.2076in
]%
{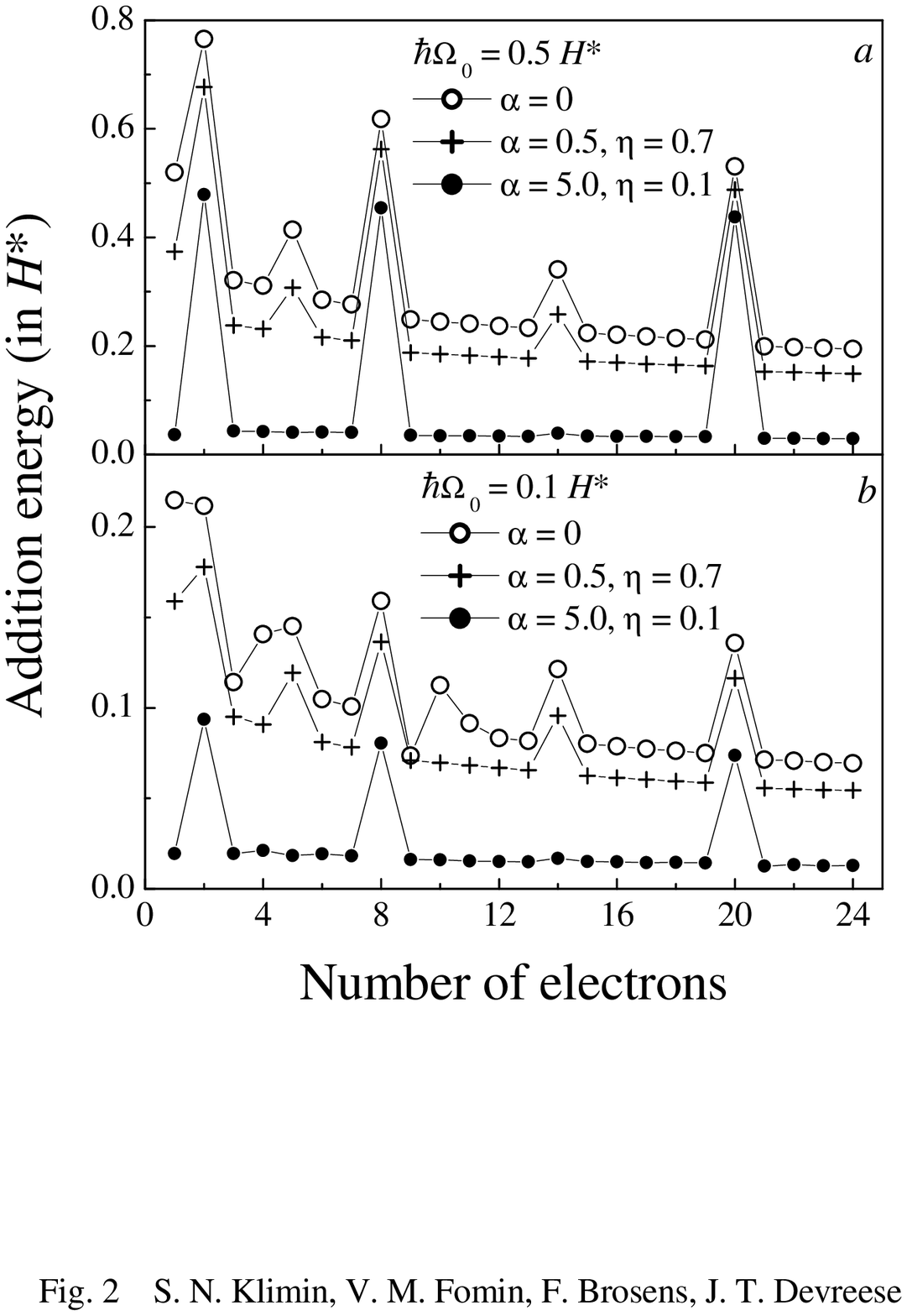}%
\end{center}
\end{figure}

\newpage%

\begin{figure}
[b]
\begin{center}
\includegraphics[
height=5.5512in,
width=3.4212in
]%
{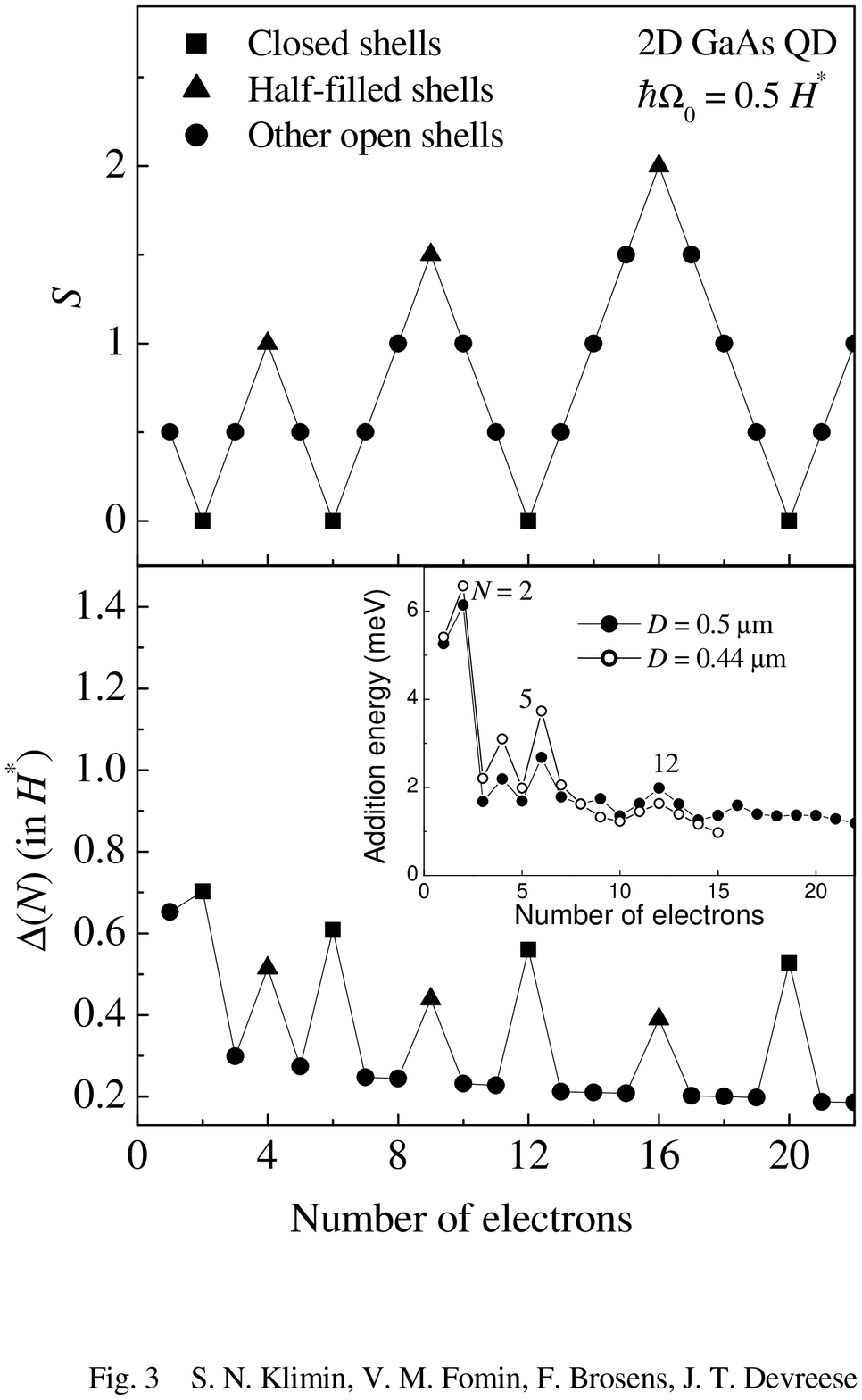}%
\end{center}
\end{figure}

\newpage%

\begin{figure}
[b]
\begin{center}
\includegraphics[
height=4.7893in,
width=4.1589in
]%
{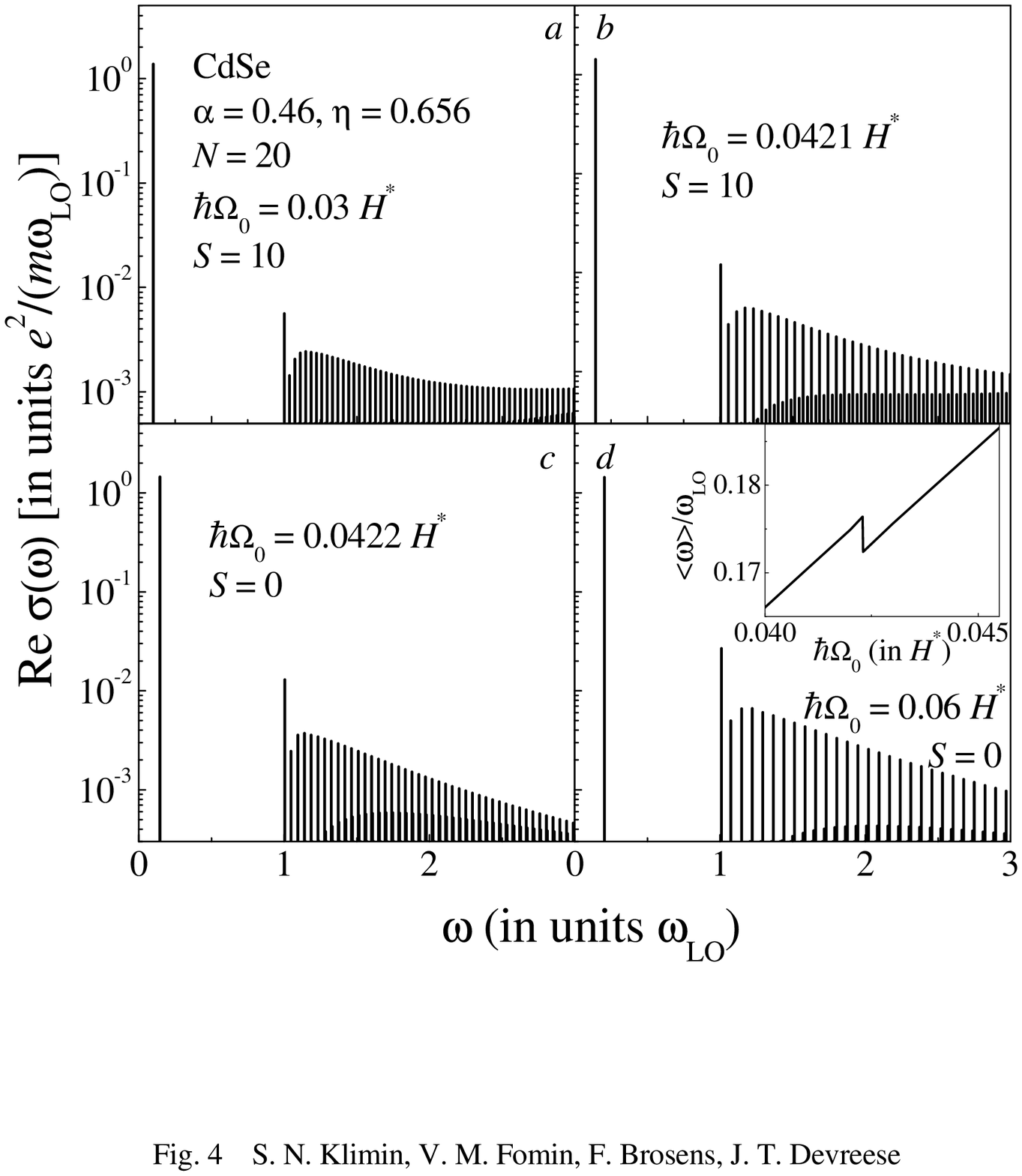}%
\end{center}
\end{figure}

\newpage%

\begin{figure}
[b]
\begin{center}
\includegraphics[
height=4.2774in,
width=3.8761in
]%
{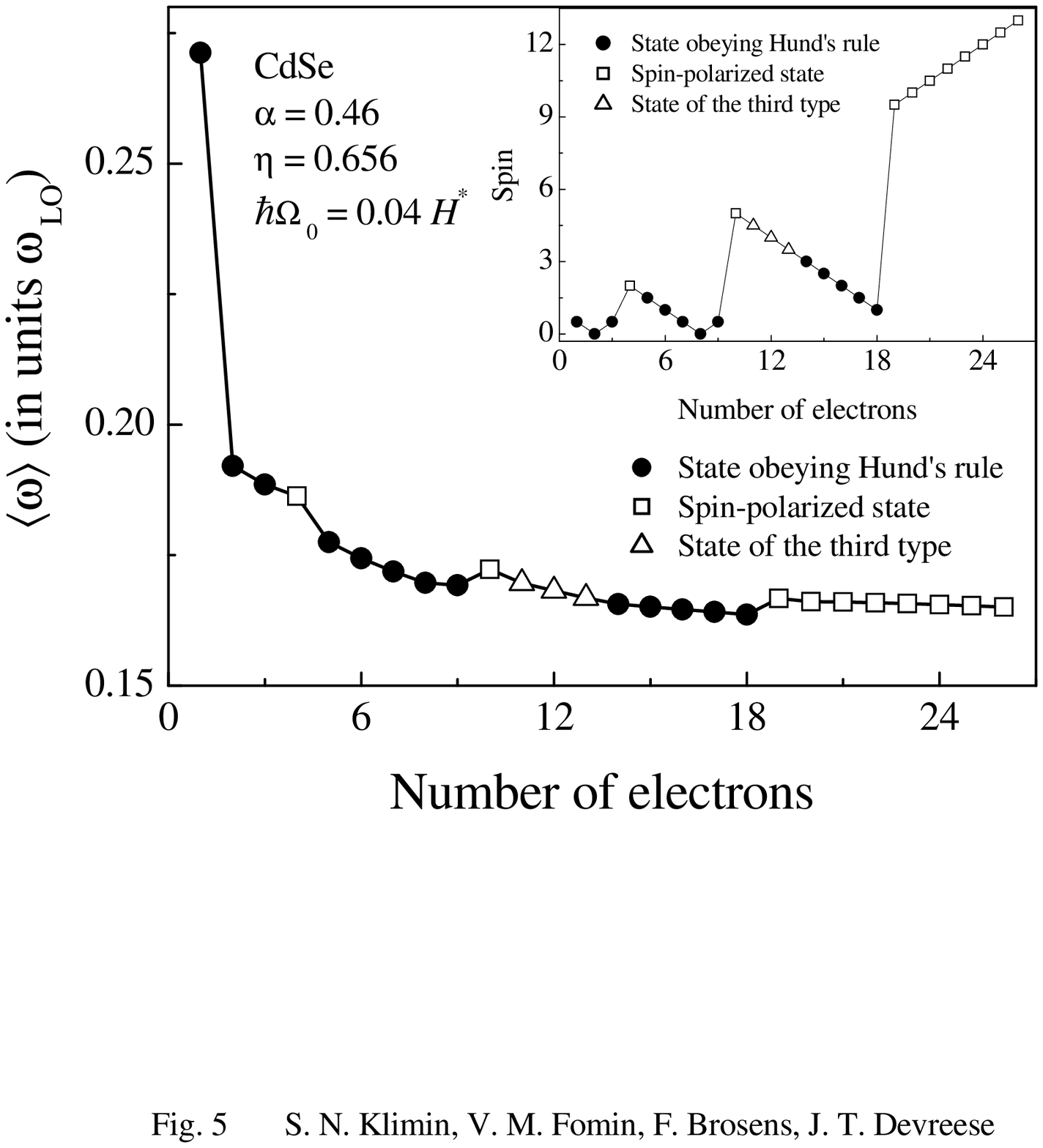}%
\end{center}
\end{figure}

\newpage%

\begin{figure}
[b]
\begin{center}
\includegraphics[
height=5.7527in,
width=4.0395in
]%
{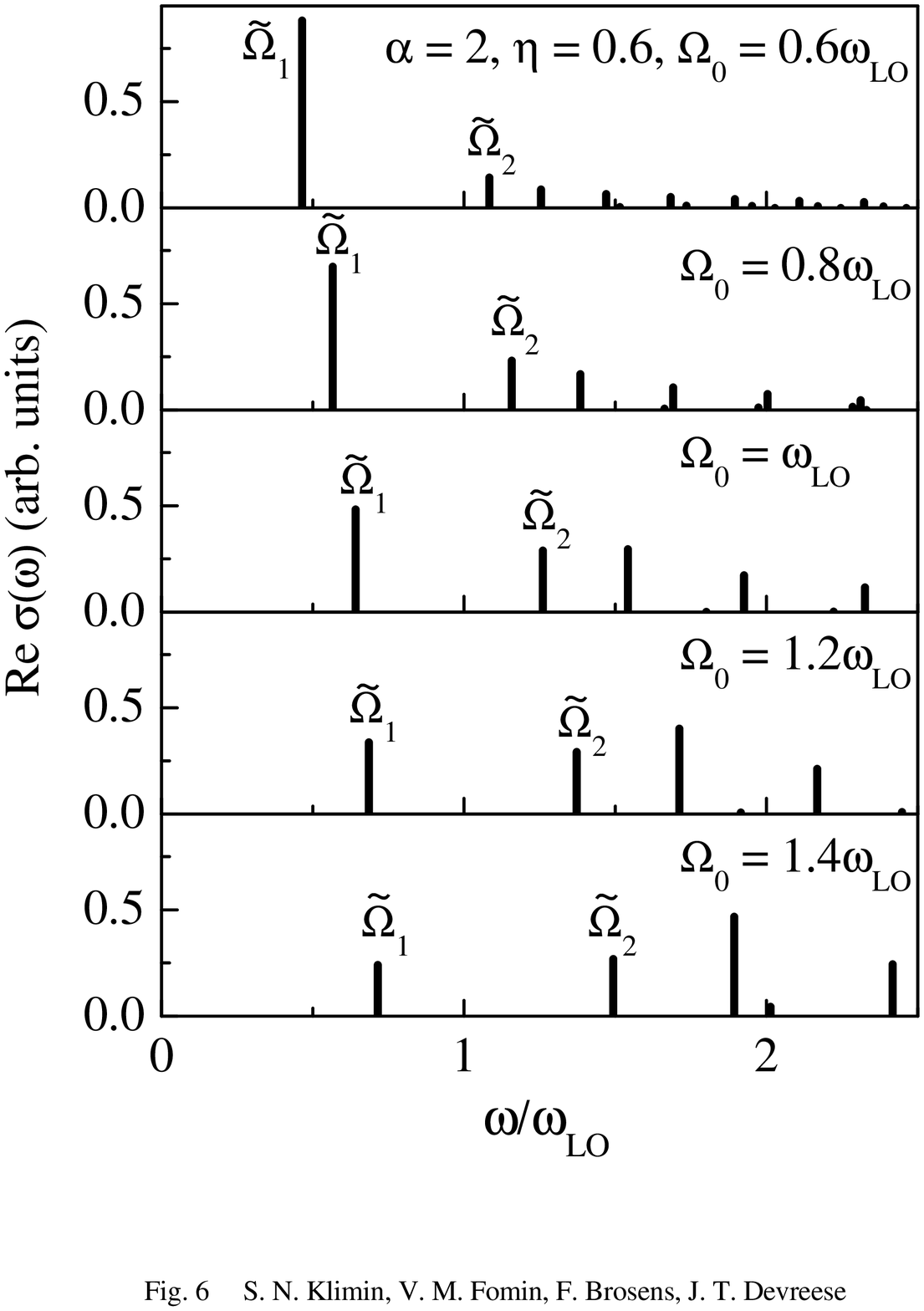}%
\end{center}
\end{figure}

\newpage%

\begin{figure}
[b]
\begin{center}
\includegraphics[
height=3.5172in,
width=4.3785in
]%
{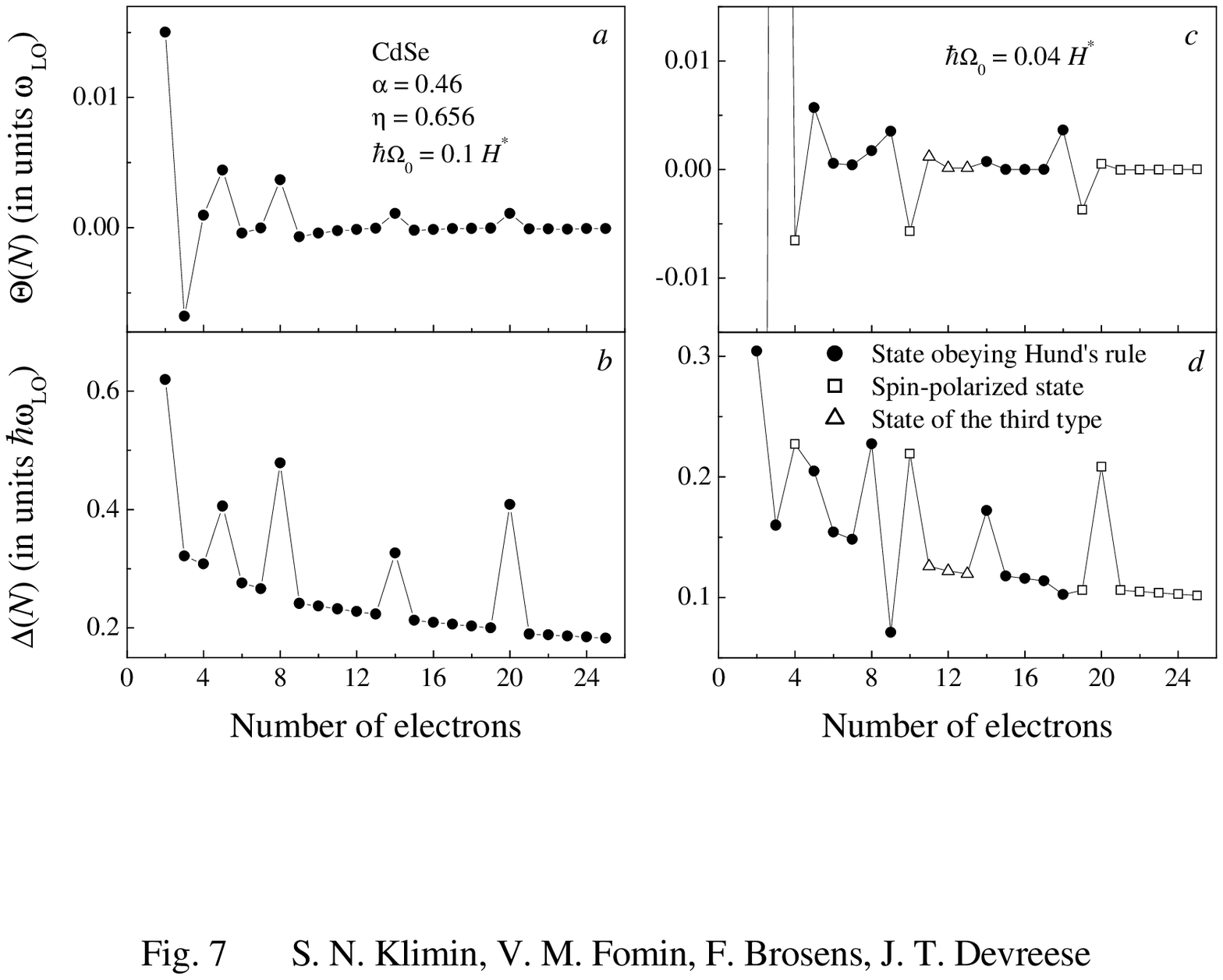}%
\end{center}
\end{figure}

\newpage%

\begin{figure}
[b]
\begin{center}
\includegraphics[
height=3.9064in,
width=4.369in
]%
{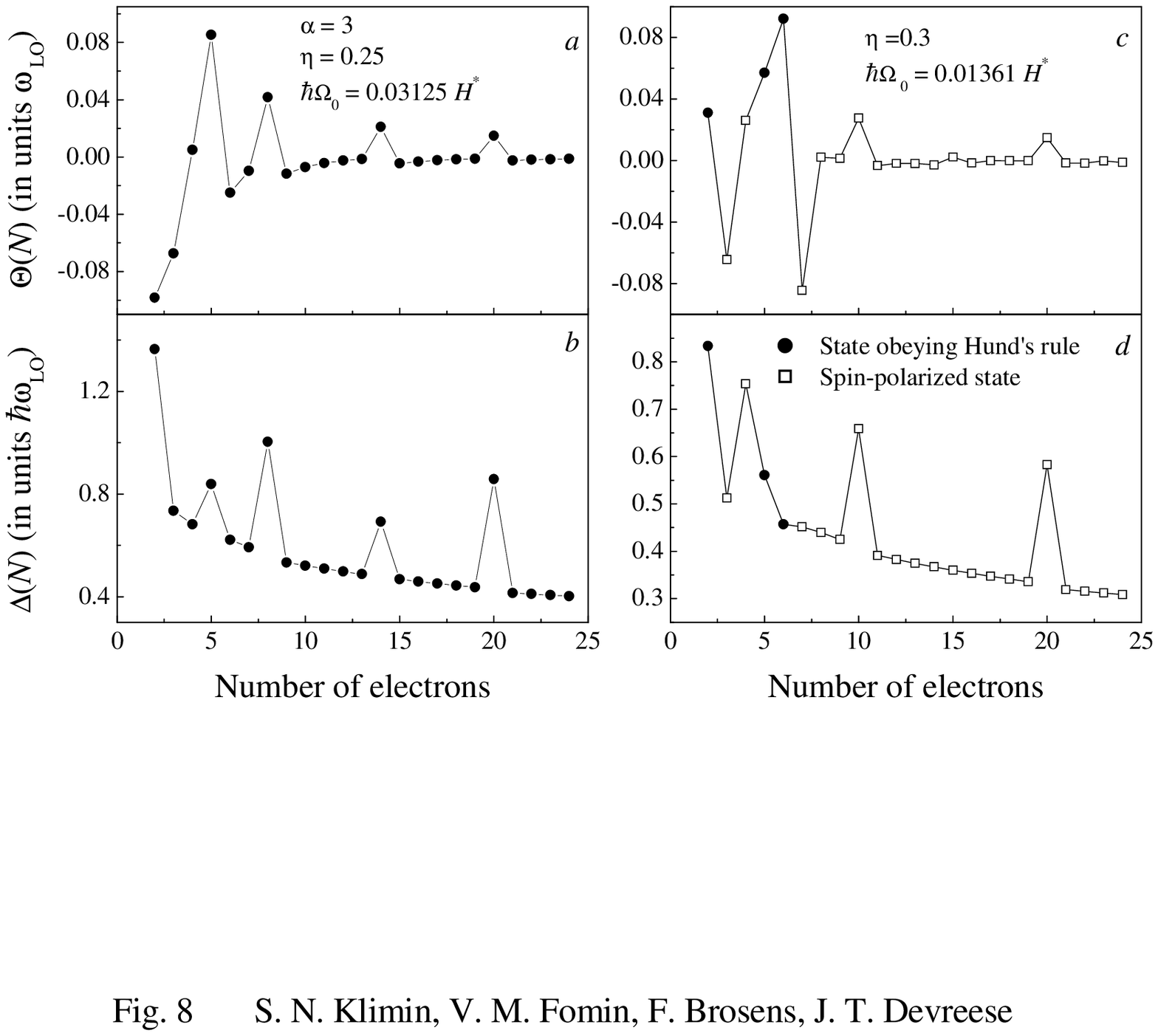}%
\end{center}
\end{figure}

\newpage%

\begin{figure}
[b]
\begin{center}
\includegraphics[
height=5.7207in,
width=4.1909in
]%
{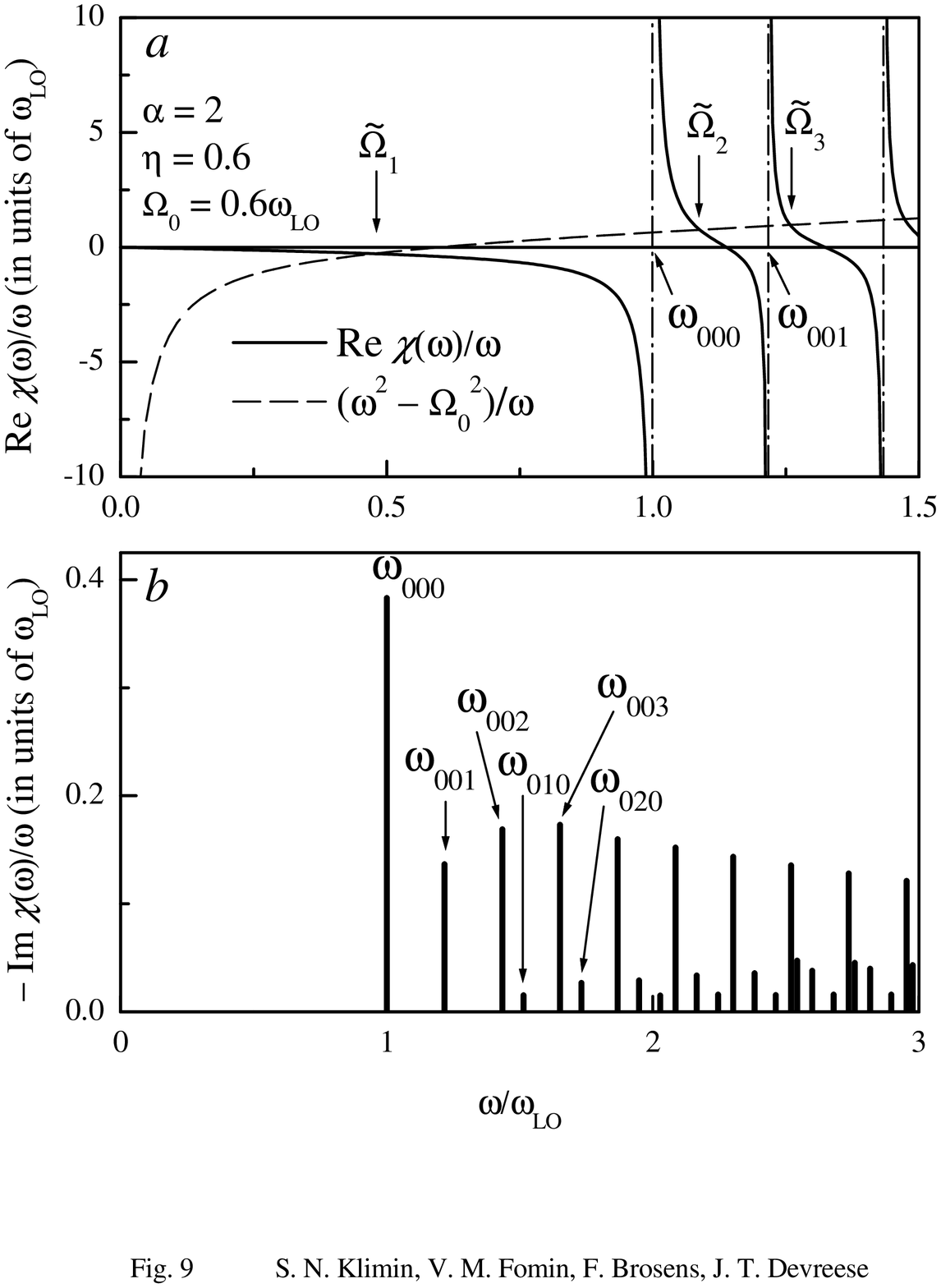}%
\end{center}
\end{figure}

\begin{thebibliography}{**}
\bibitem {alpha}Future technological progress might allow one to fabricate
quantum dots of materials with relatively large values of $\alpha$.
\end{thebibliography}
\end{document}